\documentclass{article}

\usepackage{arxiv}

\usepackage[utf8]{inputenc} 
\usepackage[T1]{fontenc}    
\usepackage{amsmath}
\usepackage{hyperref}       
\usepackage{url}            
\usepackage{booktabs}       
\usepackage{amsfonts}       
\usepackage{nicefrac}       
\usepackage{microtype}      
\usepackage{breqn}
\usepackage{graphicx}
\graphicspath{ {./images/} }
\usepackage{xcolor}
\usepackage{cite}
\usepackage{subcaption}
\usepackage{enumitem}

\title{On the traversable Yukawa-Casimir wormholes}

\author{
 P. H. F. Oliveira \\
  Departamento de Física, Universidade Federal do Ceará\\
  Campus do Pici, 60455-760, Fortaleza, Ceará, Brazil \\
  \href{mailto:pedroooliveira@fisica.ufc.br}{pedrooliveira@fisica.ufc.br} \\
  \And
 G. Alencar \\
  Departamento de Física, Universidade Federal do Ceará\\
  Campus do Pici, 60455-760, Fortaleza, Ceará, Brazil \\
\href{mailto:geova@fisica.ufc.br}{geova@fisica.ufc.br} \\
    \And
 I. C. Jardim \\
  Departamento de Física, Universidade Regional do Cariri\\
  Campus Crajubar, 63040-000, Juazeiro do Norte, Ceará, Brazil \\
  \href{mailto:ivan.jardim@urca.br}{ivan.jardim@urca.br} \\
  \And
  R. R. Landim \\
  Departamento de Física, Universidade Federal do Ceará\\
  Campus do Pici, 60455-760, Fortaleza, Ceará, Brazil \\
  \href{mailto:renan@fisica.ufc.br}{renan@fisica.ufc.br} \\
}

\begin{document}
\maketitle
\begin{abstract}
Wormholes (WH) require negative energy, and therefore an exotic matter source. Since Casimir energy is negative, it has been speculated as a good candidate to source that objects a long time ago. However only very recently a full solution for $D = 4$ has been found by Garattini, thus the Casimir energy can be a source of traversable WHs. In the manuscript published by the authors, we show that this solution is possible for all space-time of dimension $D > 3$. Recently, Garattini sought to analyze the effects of Yukawa-type terms on shape functions and obtained promising results. However, it assumes reasonably questionable premises when establishing a non-homogeneity in the Equation of State, when assuming a fixation between the constants that is only observed in the usual model and when considering the Zero Tidal Condition, not observed in Casimir's wormholes. In this work, we study the effects generated by Yukawa-type corrective factors on traversable wormhole solutions generated by Casimir energy without assuming such premises. We observe that, in addition to being possible to build traversable wormholes that satisfy all the necessary conditions, it is possible to obtain an adequate fixation of the constants in order to recover the standard case without a double limit. It was observed that, in the global and constant cases, it is possible to set a value in the shielding parameter that makes the wormhole generate a repulsive gravitational force. Finally, we observed that in the constant case there was a limitation for the shielding factor, something not observed in the original article.
\end{abstract}

\newpage

\section{Introduction}

Wormholes, like black holes, are predicted solutions to the Field Equations of General Relativity by imposing static and spherically symmetric symmetry on space-time metrics. The existence of black holes was investigated in numerous manuscripts after Schwarschild obtained the first solution in 1916 \cite{Schwarzschild:1916uq}, culminating in the most recent successes in photographing such cosmological objects \cite{Akiyama:2019cqa,Akiyama:2021tfw}. Meanwhile, the existence of wormholes remains unresolved \cite{Hassan:2022ibc}. Such solutions generally require sources of exotic matter, sources of negative energy, as they violate the standard energy conditions of General Relativity \cite{Shinkai:2002gv}. Such solutions, in different gravitational scenarios, have aroused recent interest and stood out as a promising area of research \cite{Gao:2016bin,Maldacena:2020sxe}.

The Casimir effect appears when we place  two plane parallel, closely spaced, uncharged, metallic plates in a vacuum. An attractive force between them appears as the zero modes of quantum field theory give rise to the energy between plates \cite{Hassan:2022ibc}. This effect was first discovered by Casimir \cite{Casimir:1948dh}. Experimental evidence of the Casimir effect is also known and was shown in Archimedes' vacuum weight experiment \cite{Avino:2019fdq}. The interesting feature of this effect is that an attractive force appears which is generated by negative energy. As such, it has been speculated as a good candidate to source those objects from a long time ago. However, only very recently has Garattini, in his manuscript \cite{Garattini:2019ivd}, shown that Casimir energy can be used solely to build Morris-Thorne wormholes in $3 + 1$ dimensions and explored the consequences of Quantum Weak Energy Condition (QWEC) in the traversability of the wormhole. Recently it was shown that Casimir energy can be used to construct traversable wormholes for spacetimes of $D > 3$ dimensions \cite{Oliveira:2021ypz}.

Motivated by his other works involving terms of the Yukawa type, Garattini proposed a series of modifications to the Casimir Wormhole with the shielding factor \cite{Garattini:2021kca}. The author, when modifying the shape function, obtained the associated energy via Einstein Field Equations (EFE). On the other hand, by setting $ \Phi '(r) = 0 $, i.e. Zero Tidal Condition (ZTC), he obtained the radial pressure. The tangential pressure was obtained via the conservation law, $\nabla_\mu T^{\mu\nu} = 0$.

In this paper we propose an alternative way to generate Yukawa-Casimir wormholes preserving the existing profile between Casimir pressure and energy, as well as looking for a complete solution without the requirement of the ZTC. This methodology allows, in our view, a more consistent generalization. This article is organized as follows: We briefly review Casimir wormholes from field equations to shape and redshift function solutions in Sec. \ref{sec:review}. In Sec. \ref{sec:ycwhs}, we construct the Yukawa-Casimir wormholes in three distinct approaches to modifying the shape function, i) Global correction, ii) Constant term correction, and iii) Variable term correction. Furthermore, we discuss the corrective effects of the first- and second-order shielding parameter.  Finally, we conclude our results in the last section.

\section{A review of Casimir Wormholes}\label{sec:review}

The Wormholes are interesting spherically symmetric solutions of Einstein's Field Equations because represent topological structures with a throat connecting two asymptotically flat regions of spacetime.
To obtain a traversable wormhole, Morris and Thorne, in 1988 \cite{Morris:1988cz}, write the geometry of the spacetime by the line element
\begin{equation}\label{eq:line}
	ds^2 = -e^{2\Phi(r)}dt^2 + \frac{dr^2}{1 - \frac{b(r)}{r}} + r^2d\Omega^2,
\end{equation}
where $d\Omega^2 = d\theta^2 + \sin^2\theta d\phi^2$,  $b(r)$ and $\Phi(r)$ are the shape and redshift functions, respectively, both functions of the coordinate $r\in[r_0,\infty)$, where $r_0 = \min[b( r)]$ is the WH throat value. In order to guarantee the wormhole traversability, Morris and Thorne, imposes that the redshift function, $\Phi(r)$, must to be regular everywhere for $r\in[r_0,\infty)$, and the shape function, $b(r)$, must to obey the following four properties, namely:
\\(i) The non-singularity condition,  $b(r)/r < 1$ for $r > r_0$; 
\\(ii) The existence of the throat, $b(r) = r_0$ at $r = r_0$; 
\\(iii) The asymptotic flat limit, $b(r)/r \rightarrow 0$ as $r \rightarrow \infty$; 
\\(iv) The flare-out condition, $b'(r)r - b(r) < 0$.

The line element (\ref{eq:line}), splits the EFE in the following set of component equations
\begin{subequations}
	\begin{eqnarray}
		\kappa\rho(r) &=& \frac{b'(r)}{r^2},\label{eq:EFE1}\\
		\kappa p_r(r) &=& \frac{2}{r}\left(1 - \frac{b(r)}{r}\right)\Phi'(r) - \frac{b(r)}{r^3},\label{eq:EFE2}\\
		\kappa p_t(r) &=& \left(1 - \frac{b(r)}{r}\right)\left[\Phi''(r) + \Phi'(r)\left(\Phi'(r) + \frac{1}{r}\right)\right] - \frac{b'(r)r - b(r)}{2r^2}\left(\Phi'(r) + \frac{1}{r}\right)\label{eq:EFE3},
	\end{eqnarray}
\end{subequations}
where $\kappa = 8\pi G$, $\rho(r)$ is the energy density, $p_r(r)$ is the radial pressure, and $p_t(r)$ is the lateral pressure. The set of equations (\ref{eq:EFE1})-(\ref{eq:EFE3}) are over-determined, i.e., the lateral pressure could be determined by the energy density and the radial pressure. Despite the Eq. (\ref{eq:EFE3}) can be used in this way, it is easier to use the  conservation of the stress-energy tensor equation   
\begin{equation}\label{eq:DT}
	p'_r(r) = \frac{2}{r}\left[p_t(r) - p_r(r)\right] - \left[\rho(r) + p_r(r)\right]\Phi'(r).
\end{equation}
The application of the flare-out condition in the throat implies that $b'(r_{0}) < 0$, which by the equation (\ref{eq:EFE1}), leads to $\rho(r_{0}) <0$. This result show us that, at last in the throat, an exotic type of matter is necessary to build a traversable wormhole. 

Due the fact that exotic matter is necessary to construct Morris-Thorne wormhole, in 2019 \cite{Garattini:2019ivd},  Garanttini showed that, by promoting the separation distance between the plates to coordinate status, the Casimir energy and pressure
\begin{eqnarray}
	\rho(r) &=& -\frac{\hbar c\pi^2}{720r^4}, \label{eq:energy}
\\ p_{r}(r) &=& \omega \rho(r) = 3 \rho(r), \label{EoS}      
\end{eqnarray}
are able to obtain a Morris-Thorne type traversable wormhole (WH) solution.
Thus, using Casimir energy (\ref{eq:energy}) and the Equation of State (EoS) (\ref{EoS}), Garattini obtained that
\begin{subequations}
	\begin{eqnarray}
		b(r) &=& r_0 - \frac{r_1^2}{r_0} + \frac{r_1^2}{r},\label{eq:shapecerta}\\
		\Phi(r) &=& -\frac{1}{2}\left[(\omega r_0^2 - r_1^2)\frac{\ln(r_0r + r_1^2)}{r_0^2 + r_1^2}\right] + (1-\omega)\ln(r) + (\omega r_1^2 - r_0^2)\frac{\omega r_1^2\ln(r - r_0)}{r_0^2 + r_1^2}, \label{redshift}
	\end{eqnarray}
\end{subequations}
where $r_1^2 = \frac{\pi^3l_P^2}{90}$. In order to ensure that the redshift function is finite everywhere, the solution (\ref{redshift}) imposes the following constraint $
	r_1^2 = \frac{r_0^2}{\omega} = \frac{r_0^2}{3}$.
This fixation allow us to write the shape and the redshift functions as 
\begin{subequations}
	\begin{eqnarray}
		b(r) &=& \frac{2r_0}{3} + \frac{r_0^2}{3r}\label{eq:shape3},\\
		\Phi(r) &=& \ln\left(\frac{3r}{3r + r_0}\right)\label{eq:redcasimir}.
	\end{eqnarray}
\end{subequations}
The above results satisfy the four condition for the shape function listed before, and the non-divergence of the redshift function. 
Finally, it is important to note that given the shape function, it is possible to visualize the wormhole through an incorporation procedure, given by the following integral \cite{Morris:1988cz}
\begin{equation}\label{eq:emb}
	z(r) = \pm \int_{r_0}^{r} \frac{dr'}{\sqrt{\frac{r'}{b(r')} - 1}}.
\end{equation}
The above curve, when rotated, generates the standard embedding diagram showing the throat connecting the disjoint regions.

\section{Yukawa-Casimir Wormholes}\label{sec:ycwhs}

Garattini, in his recent manuscript, seeks to propose a way to study the effect of Yukawa-type shielding terms on the shape function of Casimir wormholes. This study, part of Eq. (\ref{eq:shapecerta}) and imposes multiplicative terms of the type
\begin{equation}\label{factor}
	f(r) = e^{-\mu(r - r_0)},
\end{equation}
with $\mu$ having an inverse dimension of length, in three contexts: i) a global modification; ii) a modification only in the constant term and iii) a modification only in the variable term. From Eq. (\ref{eq:EFE1}) obtains the energy density responsible for generating the wormhole and, imposing the ZTF condition, obtains the radial pressure in terms of a non-homogeneous EoS, $p_r(r) = \omega(r) \rho(r)$.

The approach taken by Garattini lacks, a priori, questionable assumptions. Initially, we have that the adopted shape function not only depends on a homogeneous EoS of the radial pressure, but also on the fixation $\omega r_1^2 = r_0^2$ obtained to circumvent the existence of horizons in the redshift function at $r = r_0 $. Furthermore, $\omega = 3$ which is the characteristic solution of the Casimir field must be modified by a non-homogeneous EoS, as pointed out by the author. A consistent way to maintain the Casimir energy profile is from the following derivative relation \cite{Alnes:2006pa}
\begin{equation}\label{eq:eosreal}
	p_r(r) = - \frac{d}{dr}[r\rho(r)].
\end{equation}

Furthermore, by taking $\mu = 0$, the Yukawa-Casimir wormholes must, by construction, return to the Casimir wormholes for every point in spacetime, with $r \in [r_0,\infty)$, and not only at the throat, as obtained by Garattini. Finally, a wormhole that generalizes Casimir's should not lead to a ZTF condition.

Based on this, we aim to propose a new way of generating Yukawa-Casimir wormholes, starting from the shape function (\ref{eq:shapecerta}), proposing the same corrections and obtaining the energy, radial pressure, tangential pressure, redshift function and the true relationship between $r_1$ and $r_0$ for each case.

\subsection{The Global correction}
The first modification made by Garattini, was a global modification in the shape function, i.e., he multiplies the solution (\ref{eq:shapecerta}) by the factor (\ref{factor}) to obtain
\begin{equation}
	b(r) = \left(r_0 - \frac{r_1^2}{r_0} + \frac{r_1^2}{r}\right)e^{-\mu(r - r_0)}.
\end{equation}
Since the multiplicative factor is less than or equal to unity, been the equality at $r = r_0$, the modified shape function satisfies the properties (i), (ii) and (iii). To satisfy the flare-out condition, i.e., $b'(r)r -b <0$ it is necessary that
\begin{equation}
-\frac{(r_{0}^{2} -r_{1}^{2})r +2r_{1}^{2}r_{0}}{r} < \mu \left[(r_{0}^{2} -r_{1}^{2})r +r_{1}^{2}r_{0}\right],
\end{equation}
which is satisfied everywhere if $r_{1}< r_{0}$ and $\mu\geq 0$. 
So from the first EFE, Eq. (\ref{eq:EFE1}), we get the energy density compatible with the shape function
\begin{equation}\label{energy1}
	\rho(r) = -\frac{e^{-\mu(r - r_0)}}{\kappa r^4}\left[r_1^2 + \frac{\mu r}{r_0}\left(r_0^2r - r_1^2(r - r_0)\right)\right],
\end{equation}
evidently, the case where $\mu = 0$ returns the Casimir energy density, Eq. (\ref{eq:energy}), without any extra assumptions. On the other hand, in the throat, $r = r_0$, we get a modification,
\begin{equation}
	\rho(r_0) = -\frac{1}{\kappa r_0^2}\left[\mu r_0 + \frac{r_1^2}{r_0^2}\right]\label{rhor0},
\end{equation}
which is already expected, after all the energy was obtained through the derivation of the shape function.
Since $\rho(r)$ has changed, it is expected that the EoS defining the radial pressure will also change. In order to preserve the similarity character with the Casimir energy, whose radial pressure is given by Eq. \eqref{eq:eosreal}, then the relationship between $\rho$ and $p_r$ becomes inhomogeneous, $p_r(r) = \omega(r)\rho(r)$, with
\begin{equation}\label{state1}
	\omega(r) = \frac{\left[(r_0^2 - r_1^2)r^3 + r_0r_1^2r^2\right]\mu^2 + \left[(r_0^2 - r_1^2)r^2 + 3r_0r_1^2r\right]\mu + 3r_1^2r_0}{\left[(r_0^2 - r_1^2)r^2 + r_0r_1^2r\right]\mu + r_1^2r_0},
\end{equation}
which has the following properties
\begin{eqnarray}
		\omega(r_0) &=& \frac{r_0^4\mu^2 + (r_0^3 + 2r_1^2r_0)\mu + 3r_1^2}{\mu r_0^3 + r_1^2}, \label{wr0}\\
		\lim\limits_{r\rightarrow\infty}\omega(r) &\rightarrow& \infty, \;\; \mbox{for}\; \mu \neq 0,\\
		\lim\limits_{\mu\rightarrow0} \omega(r) &=& 3, 
\end{eqnarray}
returning the usual case. Although the threshold $r\rightarrow\infty$ causes $\omega$ to diverge, radial pressure has a general protection factor guaranteeing the finitude of pressure. Contrary to Garattini's proposal, it is not necessary to impose a double limit to return to the standard case.

\begin{figure}[!h]
	\centering
	\begin{subfigure}{0.48\textwidth}
		\includegraphics[width=\textwidth]{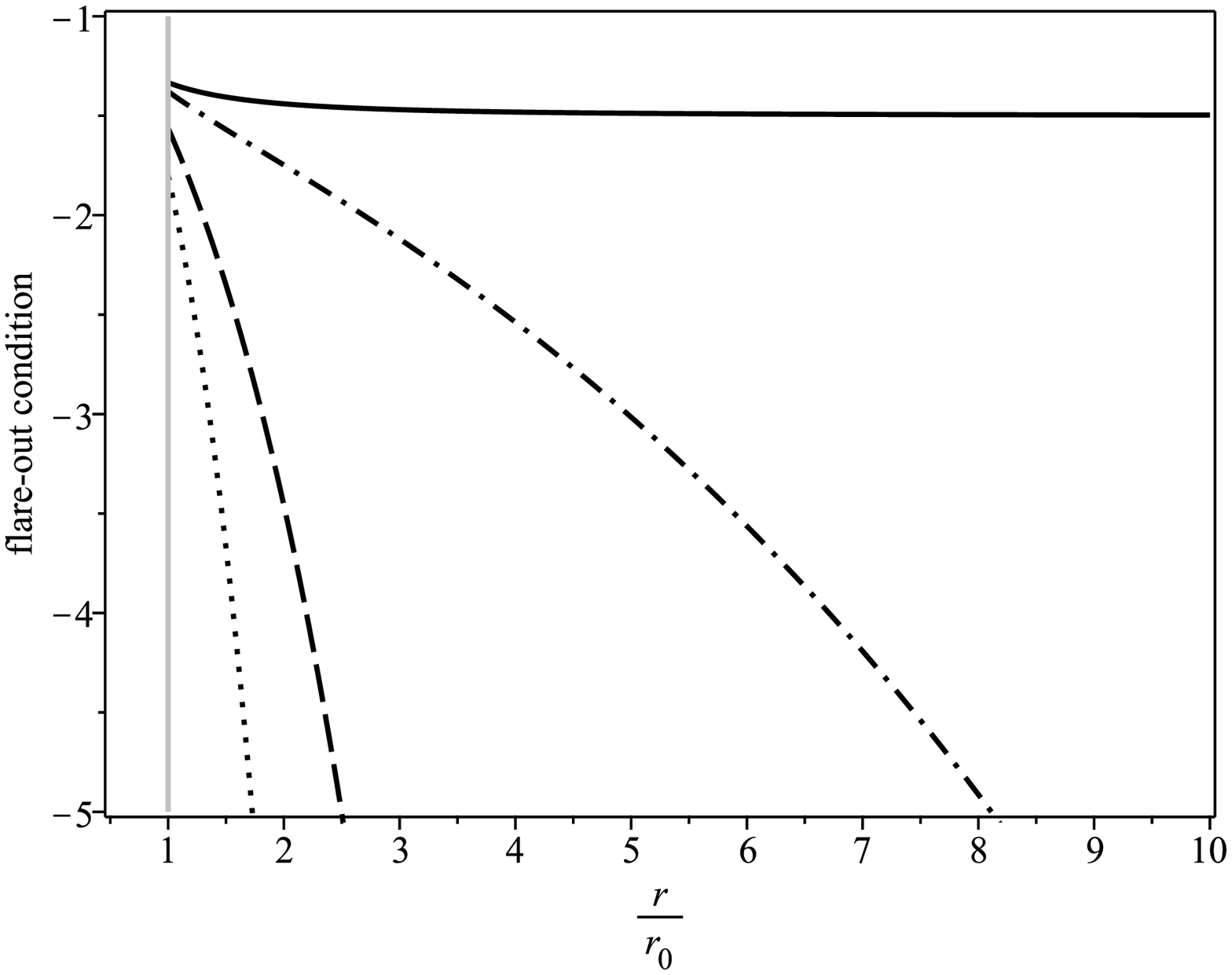}
		\caption{Flare-out condition, $[b'(r)r - b(r)]/b^2(r) <0$.}
		\label{fig:flareglobal}
	\end{subfigure}
	\hfill
	\begin{subfigure}{0.48\textwidth}
		\includegraphics[width=\textwidth]{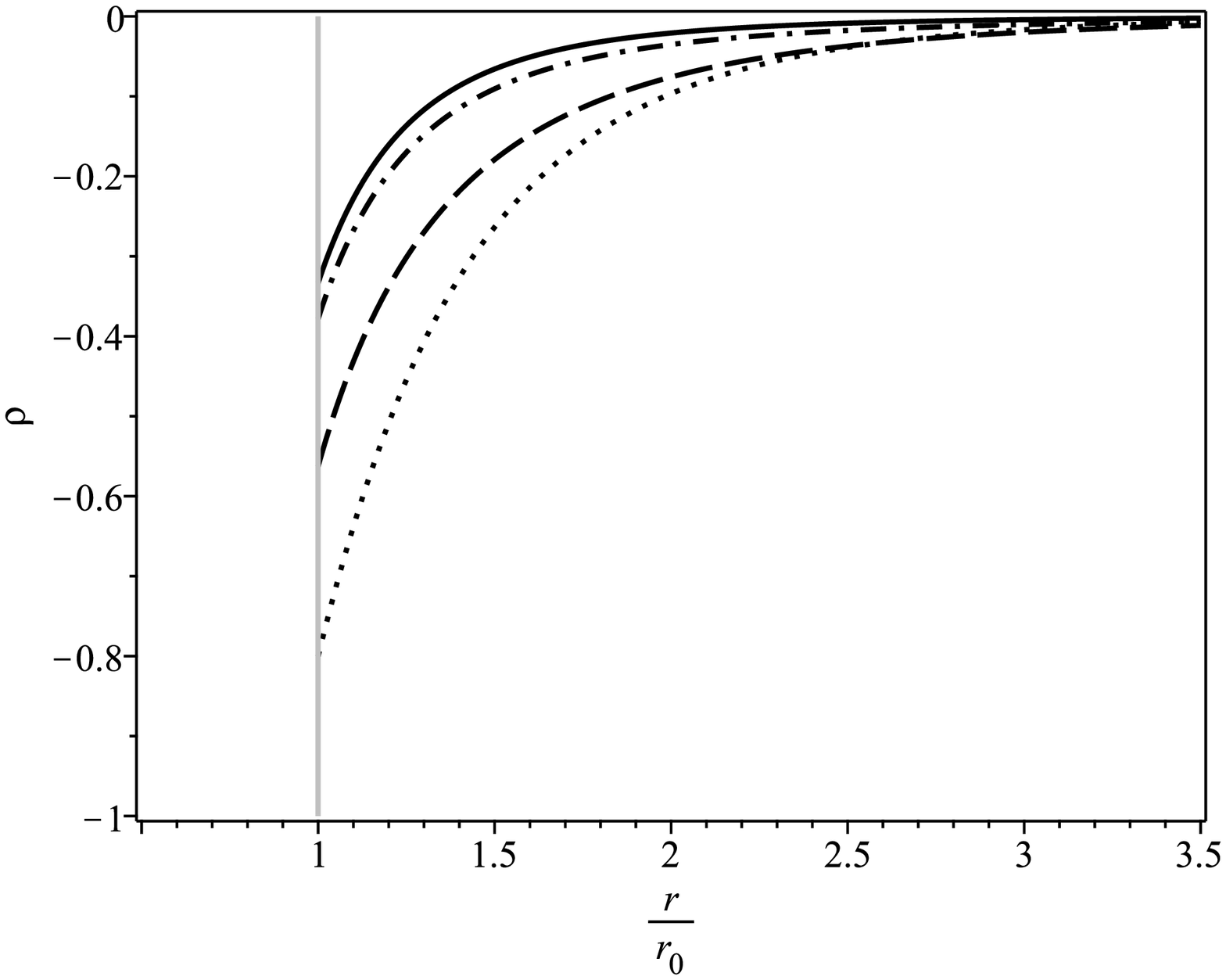}
		\caption{Energy density, $\rho(r)$.}
		\label{fig:energyglobal}
	\end{subfigure}
	\hfill
	\begin{subfigure}{0.48\textwidth}
		\includegraphics[width=\textwidth]{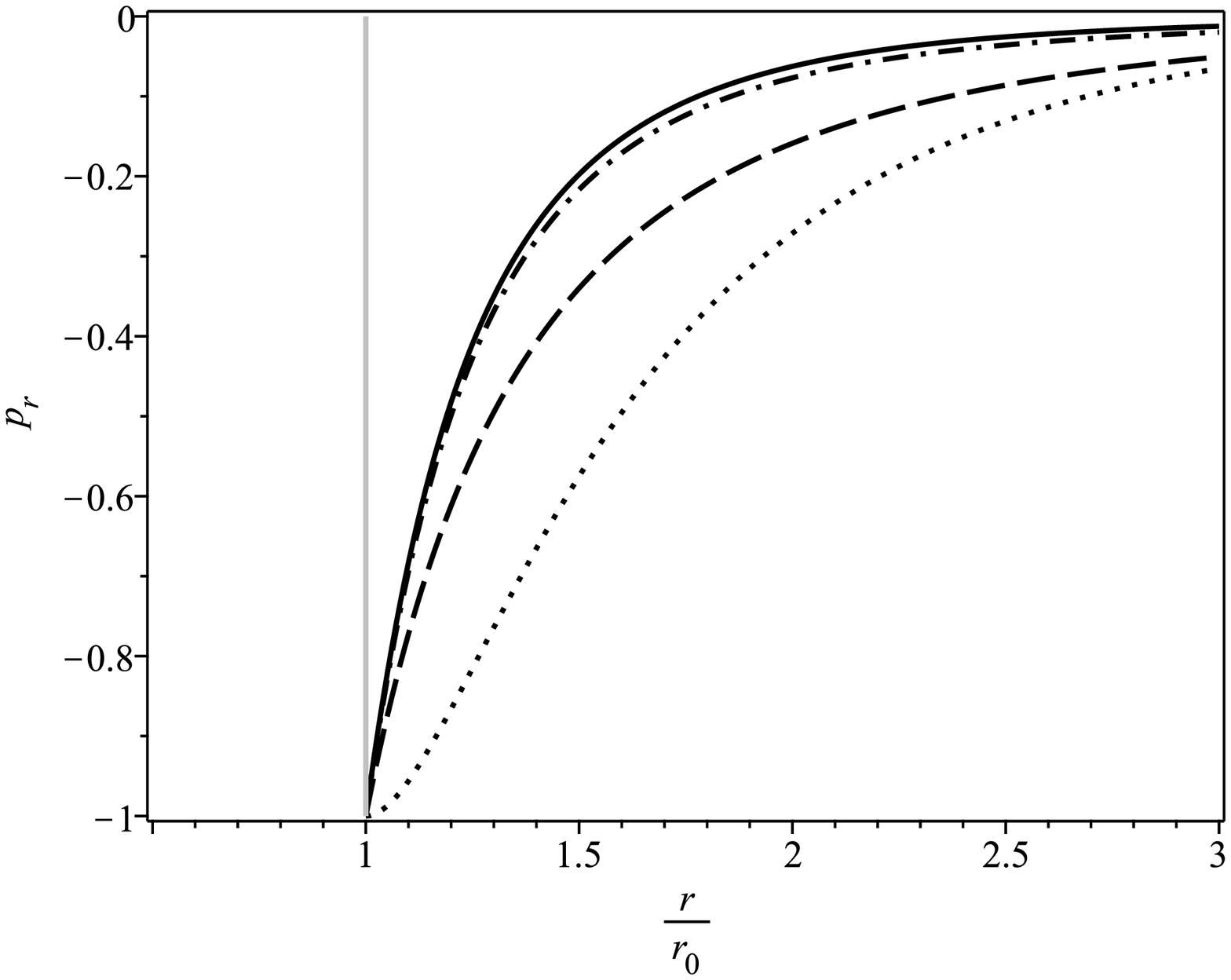}
		\caption{Radial pressure, $p_r(r)$.}
		\label{fig:radialglobal}
	\end{subfigure}
	\hfill
	\begin{subfigure}{0.48\textwidth}
		\includegraphics[width=\textwidth]{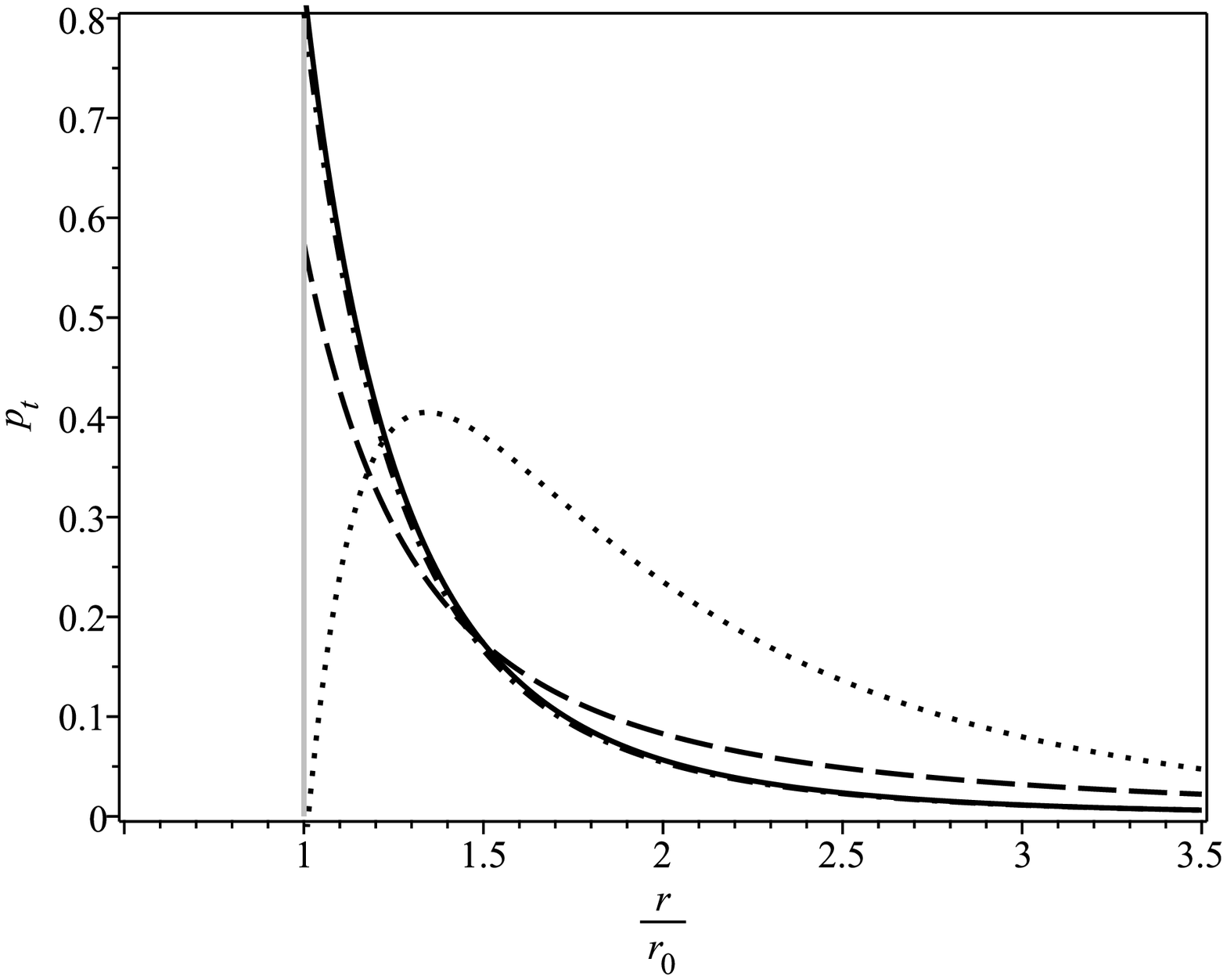}
		\caption{Tangencial pressure, $p_t(r)$.}
		\label{fig:tanglobal}
	\end{subfigure}
	
	\caption{Behavior of the flare-out condition and of the moment-energy tensor components associated with the global modification. In all graphs we have $\bar{\mu} = 0.0$ (solid line), $\bar{\mu} = 0.1$ (dashdot line), $\bar{\mu} = 0.5$ (dashed line) and $\bar{\mu} = 1.0$ (dotted line).}
	\label{fig:global}
\end{figure}

In order to avoid the divergence of the redshift function in the throat, we can apply the second EFE, Eq. (\ref{eq:EFE2}), at $r = r_{0}$. Assuming that $\Phi'(r_{0}) <\infty$, we obtain that
\begin{equation}
	kp_r(r_0) = - \frac{b(r_0)}{r_0^3} = -\frac{1}{r_{0}^{2}},
\end{equation}
where, by the use of equations (\ref{wr0}) and (\ref{rhor0}), we obtain the relation between $r_{1}$ and $r_{0}$,
\begin{equation}
	r_1^2 = \left[\frac{1 - \mu^2r_0^2 - \mu r_0}{3 + 2\mu r_0}\right]r_0^2.
\end{equation}
The above fixation removes the divergence of the redshift function in the throat, agree whit the condition $r_{1} < r_{0}$ and, for $\mu = 0$, recovers the standard result $r_0^2 = 3r_1^2$. Thus, it is possible to rewrite the energy density (\ref{energy1}) and the state function (\ref{state1}, as  
\begin{eqnarray}
	\bar{\rho}(u)  &=& -\frac{e^{-\bar{\mu}  \left(u -1 \right)} \left[\left(u^{2} -u\right)\bar{\mu}^3+\left(3u^2- u - 1\right) \bar{\mu}^{2}+\left(2 u^{2}+u -1\right)\bar{\mu} +1 \right]}{\left(2\bar{\mu} +3\right) u^{4}} \label{barrhoglobal},
\\ 	\omega(r) &=& \frac{(u^{3}-u^{2})\bar{\mu}^{4} +(3u^{3}- 3u)\bar{\mu}^{3} +(2u^{3} +4u^{2}-3u -3)\bar{\mu}^{2}  +(2u^{2} +3u -3)\bar{\mu} +3}{(u^{2} -u)\bar{\mu}^{3} +(3u^{2}- u -1)\bar{\mu}^2 +(2u^2 +u -1)\bar{\mu} + 1 },
\end{eqnarray}
where we defined the dimensionless energy density, $\bar{\rho}(u) = \kappa r_{0}^{2}\rho(u)$, radial coordinate, $u = r/r_{0}$, and mass parameter, $\bar{\mu} = \mu r_{0}$.    
Furthermore, to obtain the tangential pressure, we use conservation of energy (\ref{eq:DT}) to obtain $p_t(r)$ and make $\Omega(r) = p_t(r)/\rho(r)$, satisfying a non-homogeneous EoS. Due to the size of the equation we will not display it, just some graphs indicating its behavior.
Figure \ref{fig:flareglobal} shows the flare-out condition for some values of dimensionless parameter $\bar{\mu}$, and illustrates the fact that it is satisfied for every $\bar{\mu} >0$. In the figures \ref{fig:energyglobal} and \ref{fig:radialglobal}, we plot the dimensionless energy density (\ref{barrhoglobal}) and the radial pressure, respectively,  for some values of $\bar{\mu}$. It shows that the global modification decreases the energy and the radial pressure when $\bar{\mu}$ increases, due the existence of massive modes with negative energy. The numerical result of the tangencial pressure are plotted in fig. \ref{fig:tanglobal} for some values of $\bar{\mu}$. Due it is made only numerically the exotic behavior for $\bar{\mu} =1.0$ can not be understood.  

\begin{figure}[!h]
	\centering
	\begin{subfigure}{0.48\textwidth}
		\includegraphics[width=\textwidth]{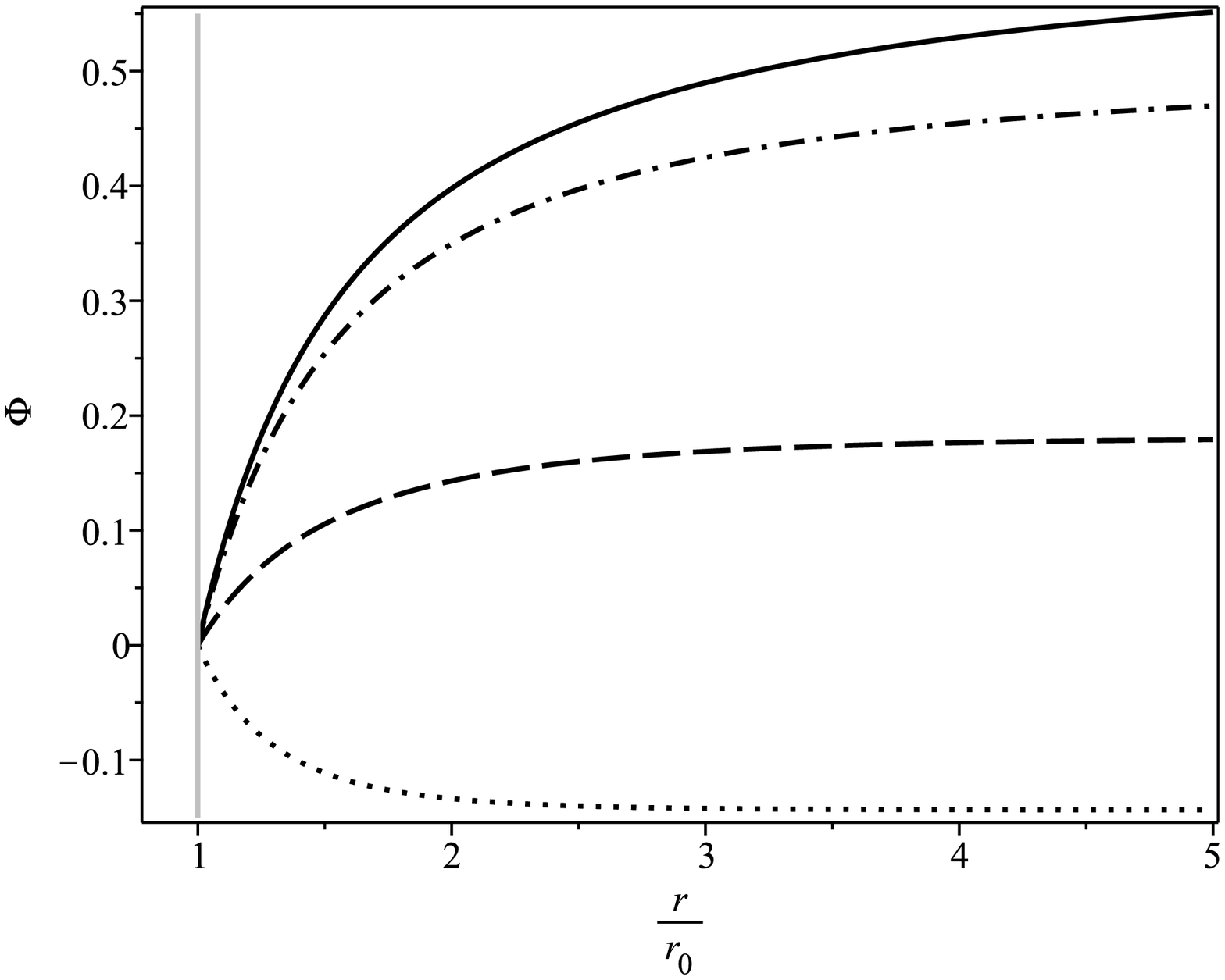}
		\caption{Redshift function, $\Phi(r)$.}
		\label{fig:redglobal}
	\end{subfigure}
	\hfill
	\begin{subfigure}{0.48\textwidth}
		\includegraphics[width=\textwidth]{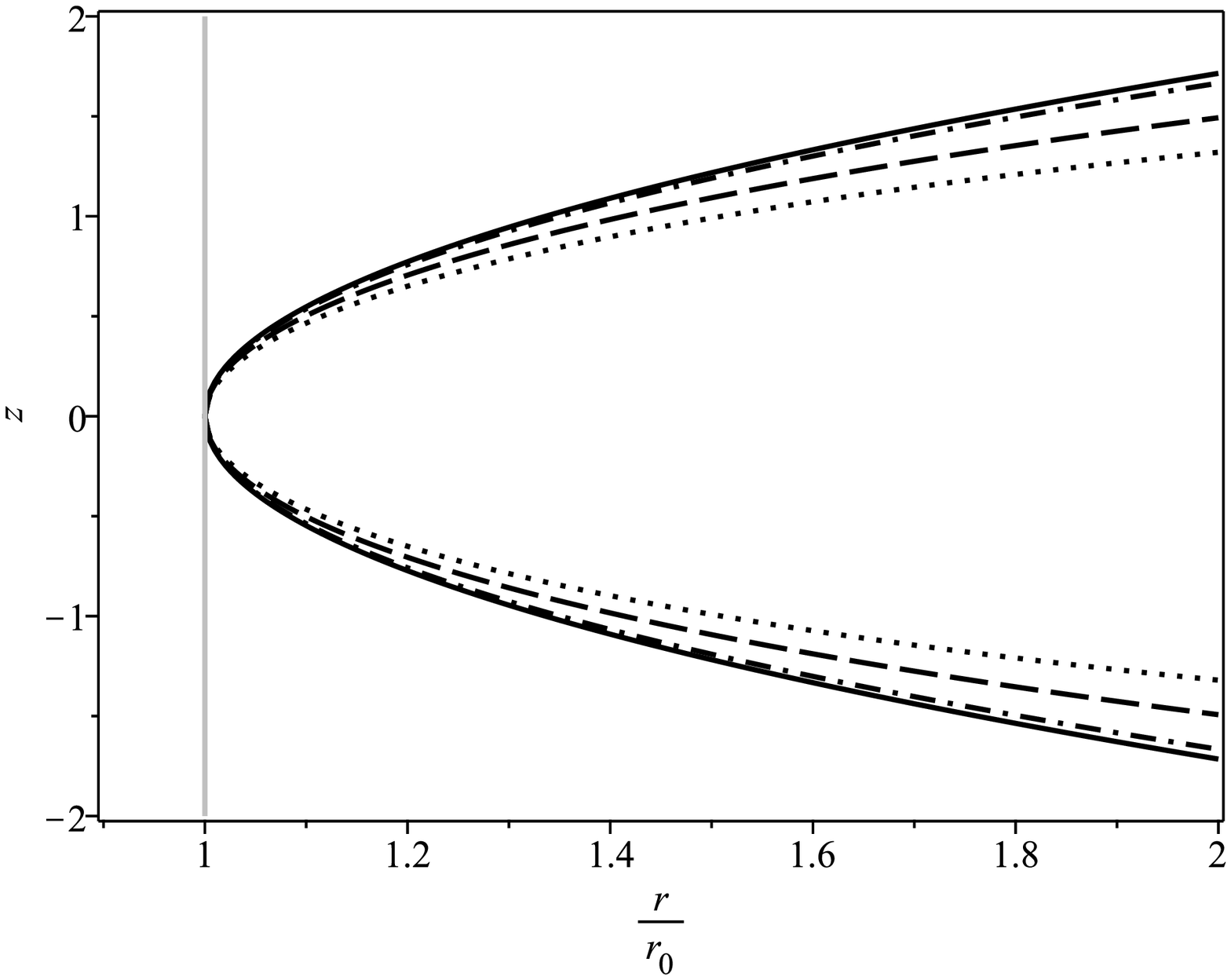}
		\caption{Embedding diagram, $z(r)$.}
		\label{fig:embglobal}
	\end{subfigure}
	
	\caption{In all graphs we have $\bar{\mu} = 0.0$ (solid line), $\bar{\mu} = 0.1$ (dashdot line), $\bar{\mu} = 0.5$ (dashed line) and $\bar{\mu} = 1.0$ (dotted line).}
	\label{fig:global2}
\end{figure}

Finally, it is possible to obtain the redshift from Eq. (\ref{eq:EFE2}) and the embedding diagram by Eq. (\ref{eq:emb}), both numerically, which are represented in Figures (\ref{fig:redglobal}) and (\ref{fig:embglobal}), respectively.

\subsection{Constant term correction}

The second type of proposed correction is of the type
\begin{equation}
	b(r) = \left(r_0 - \frac{r_1^2}{r_0}\right)e^{-\mu(r - r_0)} + \frac{r_1^2}{r}, 
\end{equation}
that is, only in the constant term. In the original case, i.e. without the Yukawa-type correction, the constant term in the shape function did not have much relevance, since most of the physical interest quantities are obtained via the derivation of $b(r)$. In this correction, however, this term should be relevant. Naturally, $b(r_0) = r_0$ and
\begin{equation}\label{eq:flarecte}
	b'(r_0) = -\frac{[r_1^2 + \mu r_0(r_0^2 - r_1^2)]}{r_0^2} < 1,
\end{equation}
provided $r_0^2 > r_1^2$, checkable later. From the first EFE we obtain the energy density that generates such a wormhole
\begin{equation}
	\rho(r) = - \frac{r_1^2}{kr^4} - \frac{\mu}{kr_0r^2}(r_0^2 - r_1^2)e^{-\mu(r - r_0)},
\end{equation}
the first term refers to the Casimir field that is retrieved if $\mu = 0$, while the second comes from the Yukawa correction. The sign of energy is preserved by the condition $r_0^2 > r_1^2$. Consequently, the radial pressure is given by
\begin{equation}
	p_r(r) = - \frac{3r_1^2}{kr^4} - \frac{(r_0^2 - r_1^2)\mu(\mu r + 1)}{kr_0r^2}e^{-\mu(r - r_0)},
\end{equation}
similarly, the first term is purely Casimir, while the second term is correction. The relationship between $\rho$ and $p_r$ is given by a non-homogeneous EoS fixed by
\begin{equation}
	\omega(r) = \frac{\mu r^2(r_0^2 - r_1^2)(\mu r + 1)e^{-\mu(r - r_0)} + 3r_1^2r_0}{\mu r^2(r_0^2 - r_1^2)e^{-\mu(r - r_0)} + r_1^2r_0},
\end{equation}
which for $\mu = 0$ returns the default value $\omega = 3$. 

\begin{figure}[!h]
	\centering
	\begin{subfigure}{0.48\textwidth}
		\includegraphics[width=\textwidth]{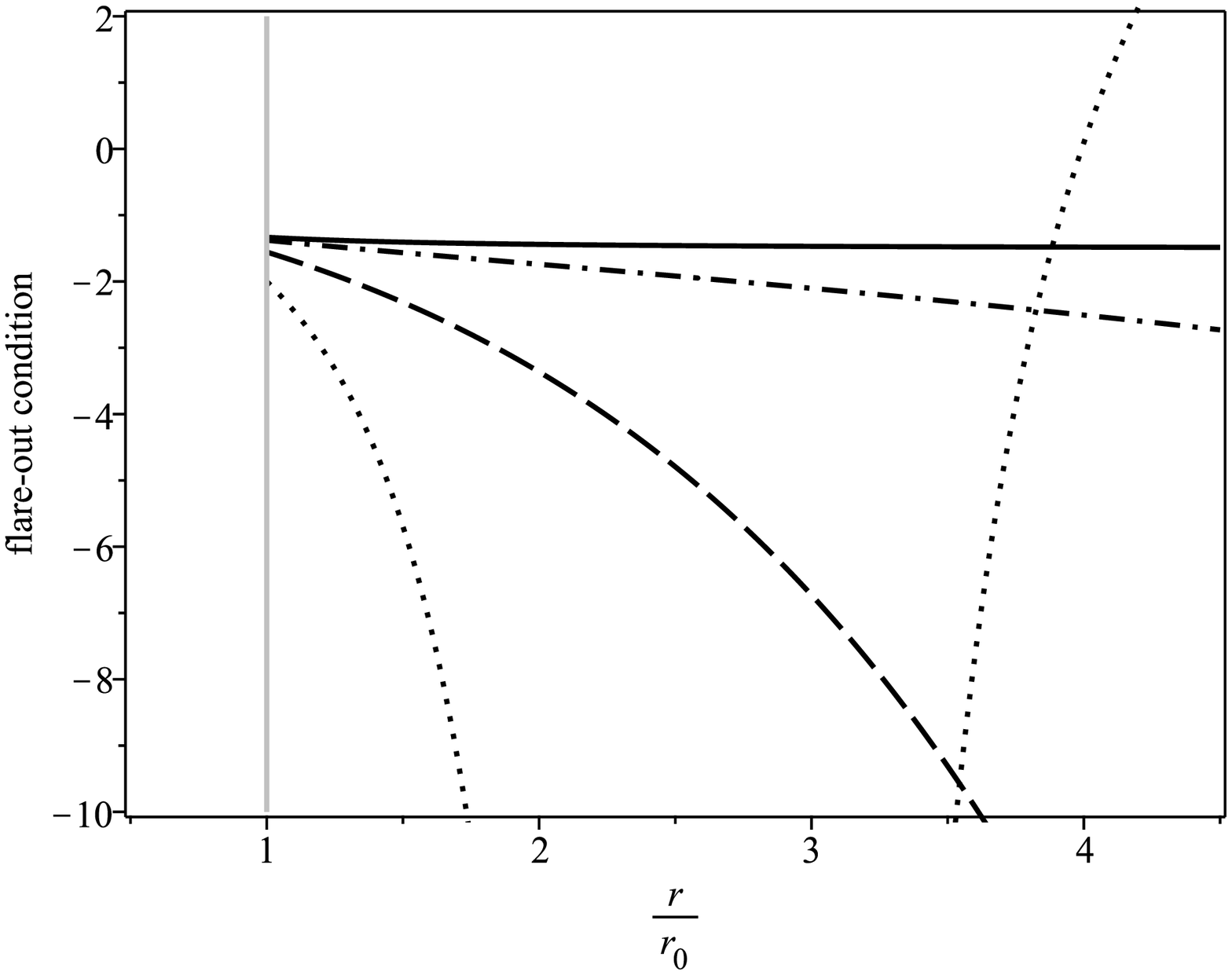}
		\caption{Flare-out condition, $[b'(r)r - b(r)]/b^2(r) < 0$.}
		\label{fig:flarecte}
	\end{subfigure}
	\hfill
	\begin{subfigure}{0.48\textwidth}
		\includegraphics[width=\textwidth]{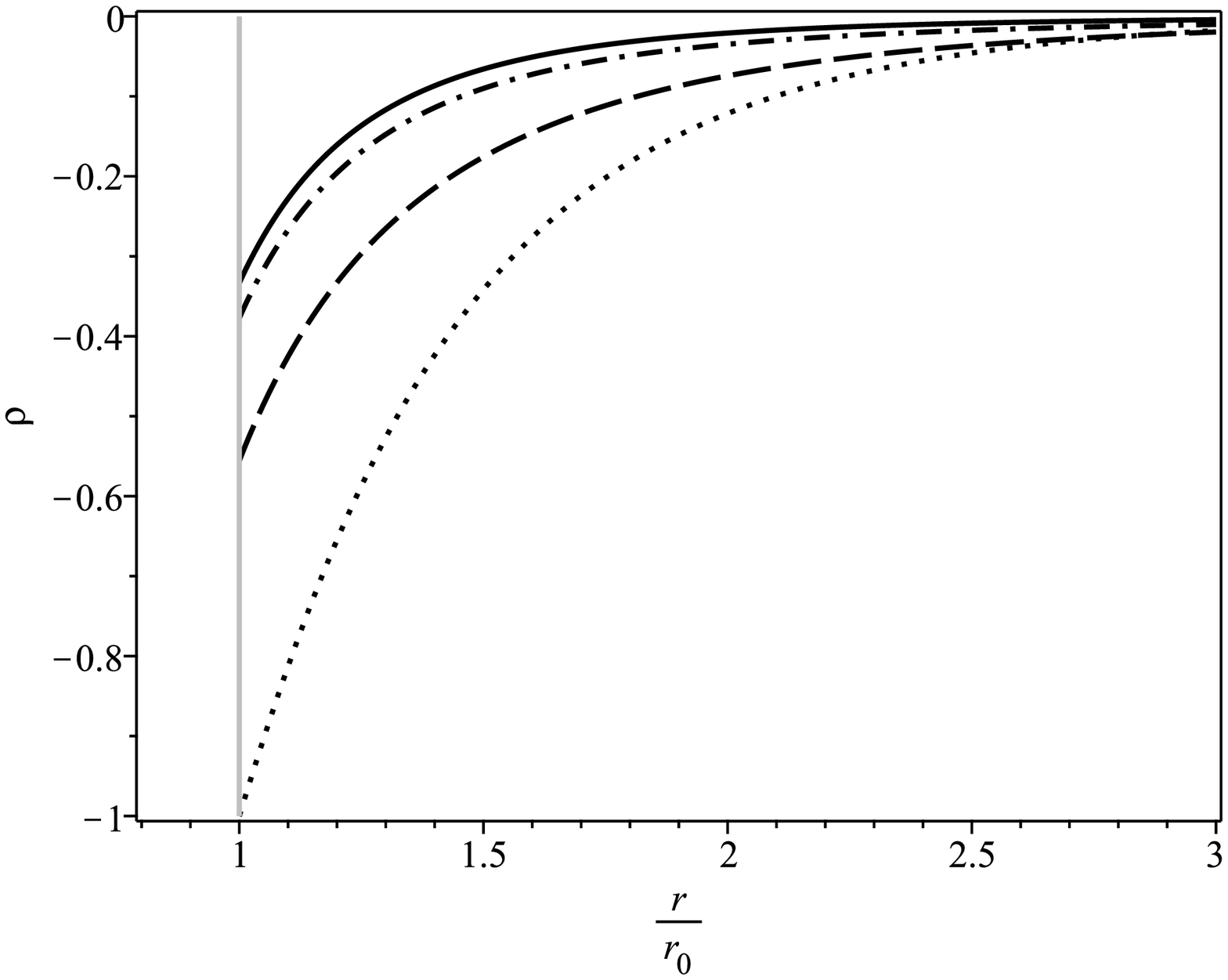}
		\caption{Energy density, $\rho(r)$.}
		\label{fig:energycte}
	\end{subfigure}
	\hfill
	\begin{subfigure}{0.48\textwidth}
		\includegraphics[width=\textwidth]{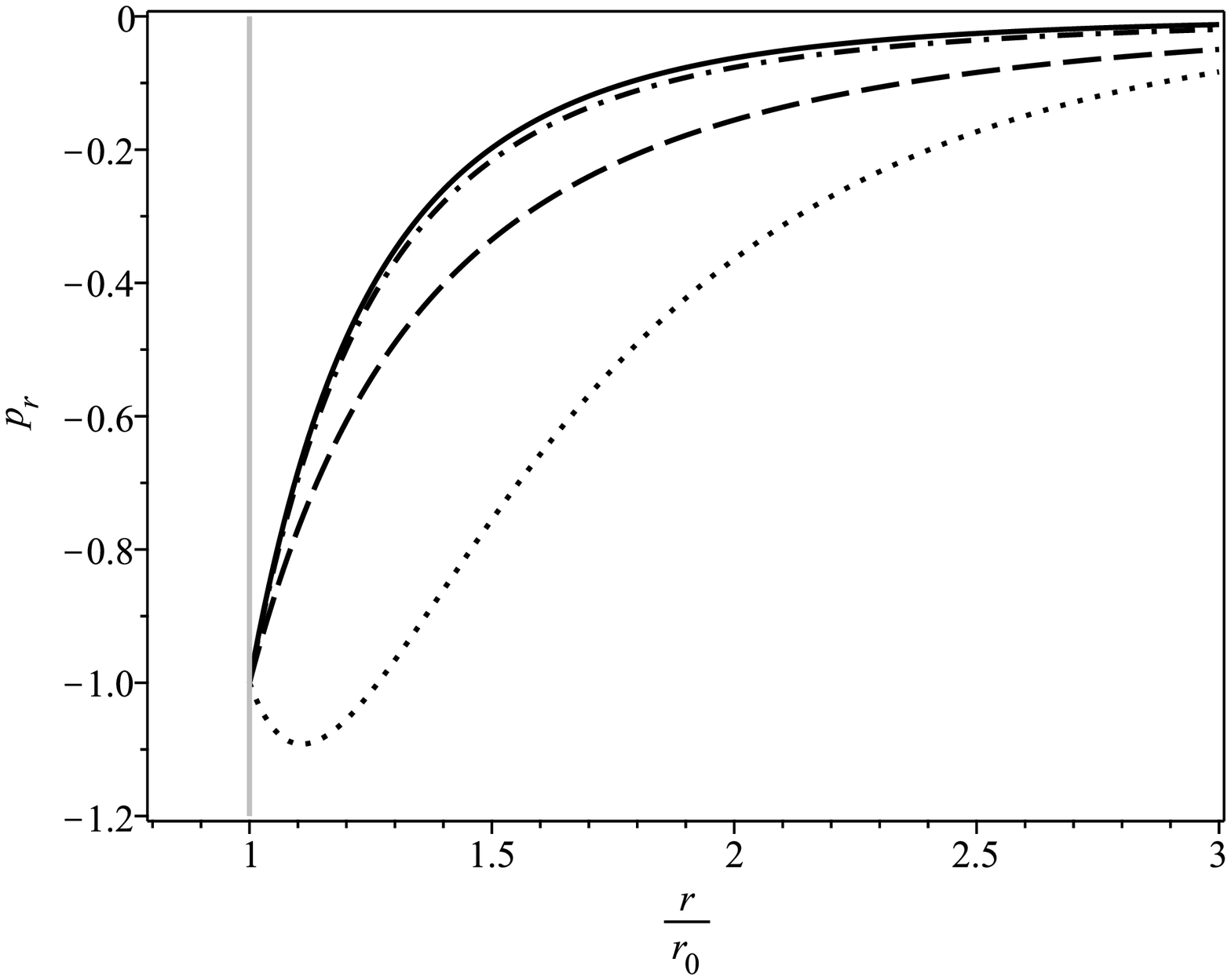}
		\caption{Radial pressure, $p_r(r)$.}
		\label{fig:radialcte}
	\end{subfigure}
	\hfill
	\begin{subfigure}{0.48\textwidth}
		\includegraphics[width=\textwidth]{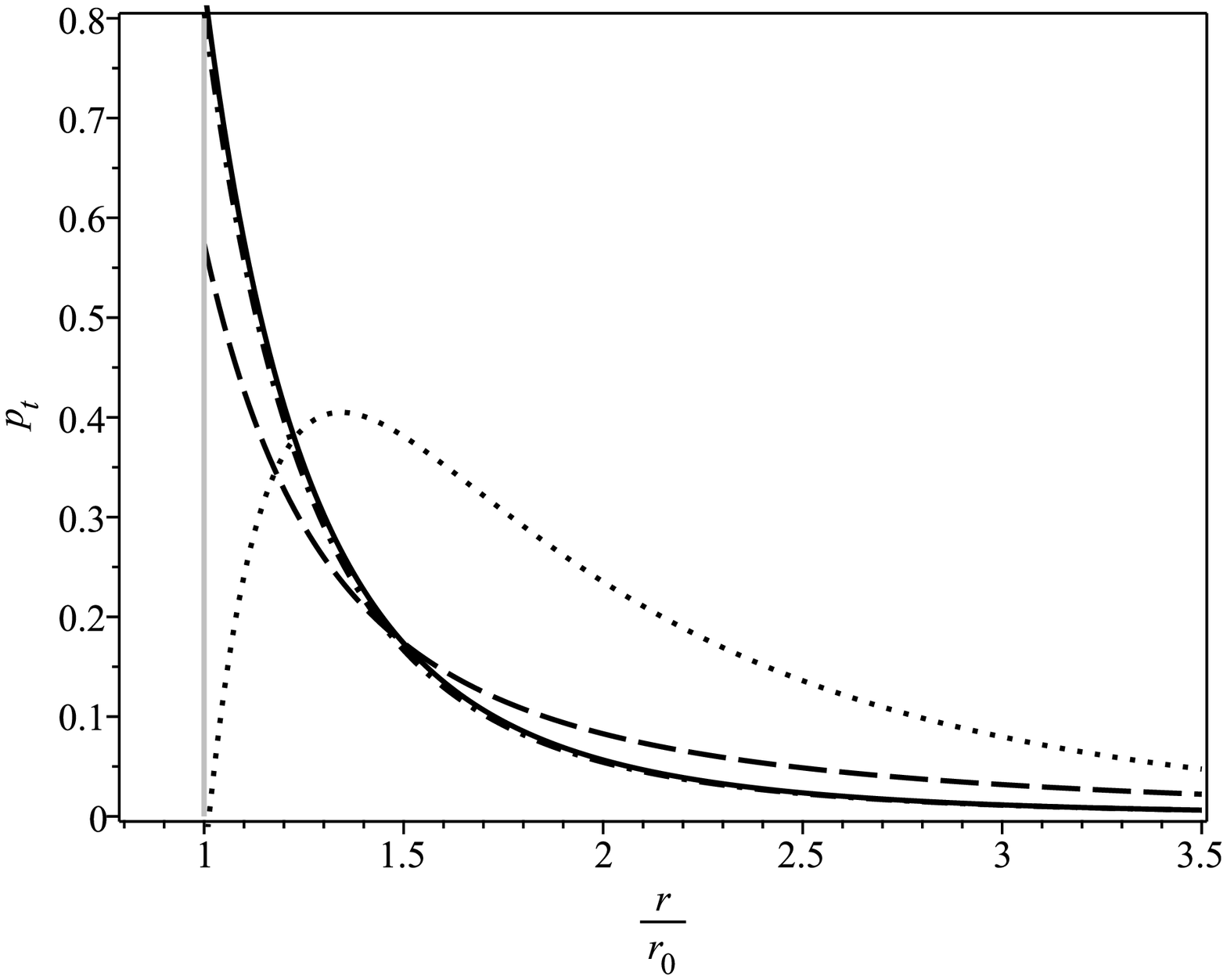}
		\caption{Tangencial pressure, $p_t(r)$.}
		\label{fig:tangcte}
	\end{subfigure}
	
	\caption{Behavior of the flare-out condition and of the moment-energy tensor components associated with the constant term correction. In all graphs we have $\mu = 0.0$ (solid line), $\mu = 0.1$ (dashdot line), $\mu = 0.5$ (dashed line) and $\mu = 1.0$ (dotted line).}
	\label{fig:cte}
\end{figure}

On the other hand, this fixation, by providing the contribution of a previously negligible term, imposes a restriction on the shielding constant. This limit is visualized in Fig. (\ref{fig:flarecte}) where the value $\mu = 1.0$ (comparable to $r_0$) violates the flare-out condition.

From the second EFE, we obtain the fixation between $r_1$, $r_0$ and $\mu$,
\begin{equation}
	r_1^2 = \left[\frac{\mu^2r_0^2 + \mu r_0 - 1}{\mu^2r_0^2 + \mu r_0 - 3}\right]r_0^2,
\end{equation}
for $\mu = 0$ we have $r_0^2 = 3r_1^2$. Furthermore, it is worth noting that $r_0 > r_1$, inequality necessary for the flare-out condition (\ref{eq:flarecte}) to be satisfied. With this fixation, we get
\begin{eqnarray}
    \bar{\rho}(u) &=& - \frac{[\bar{\mu}^2 + (1- 2u^2e^{-\bar{\mu}(u - 1)})\bar{\mu} - 1]}{u^{4}(\bar{\mu}^2 + \bar{\mu} - 3)},\\
    \bar{p}_r(u) &=& - \frac{3(\bar{\mu}^2 + \bar{\mu} - 1)}{u^{4}(\bar{\mu}^2 + \bar{\mu} - 3)} + \frac{2\bar{\mu}(\bar{\mu} u + 1)}{u^2(\bar{\mu}^2 + \bar{\mu} - 3)}e^{-\bar{\mu}(u - 1)},\\
    \Phi'(u) &=&  \frac{(-\bar{\mu}^2 - \bar{\mu} + 1) + (\bar{\mu}u^{2}(\bar{\mu} u + 1) -u^{2})e^{-\bar{\mu}(u - 1)}}{\left((\bar{\mu}^2 + \bar{\mu} - 3)u^{2} +2ue^{-\bar{\mu}(u - 1)} - (\bar{\mu}^2 + \bar{\mu} - 1)\right)}.
\end{eqnarray}
Meanwhile, the tangential pressure is obtained by the conservation of energy. Expressions, due to size, will be omitted but some graphs indicating their behavior will be displayed.

\begin{figure}[!h]
	\centering
	\begin{subfigure}{0.48\textwidth}
		\includegraphics[width=\textwidth]{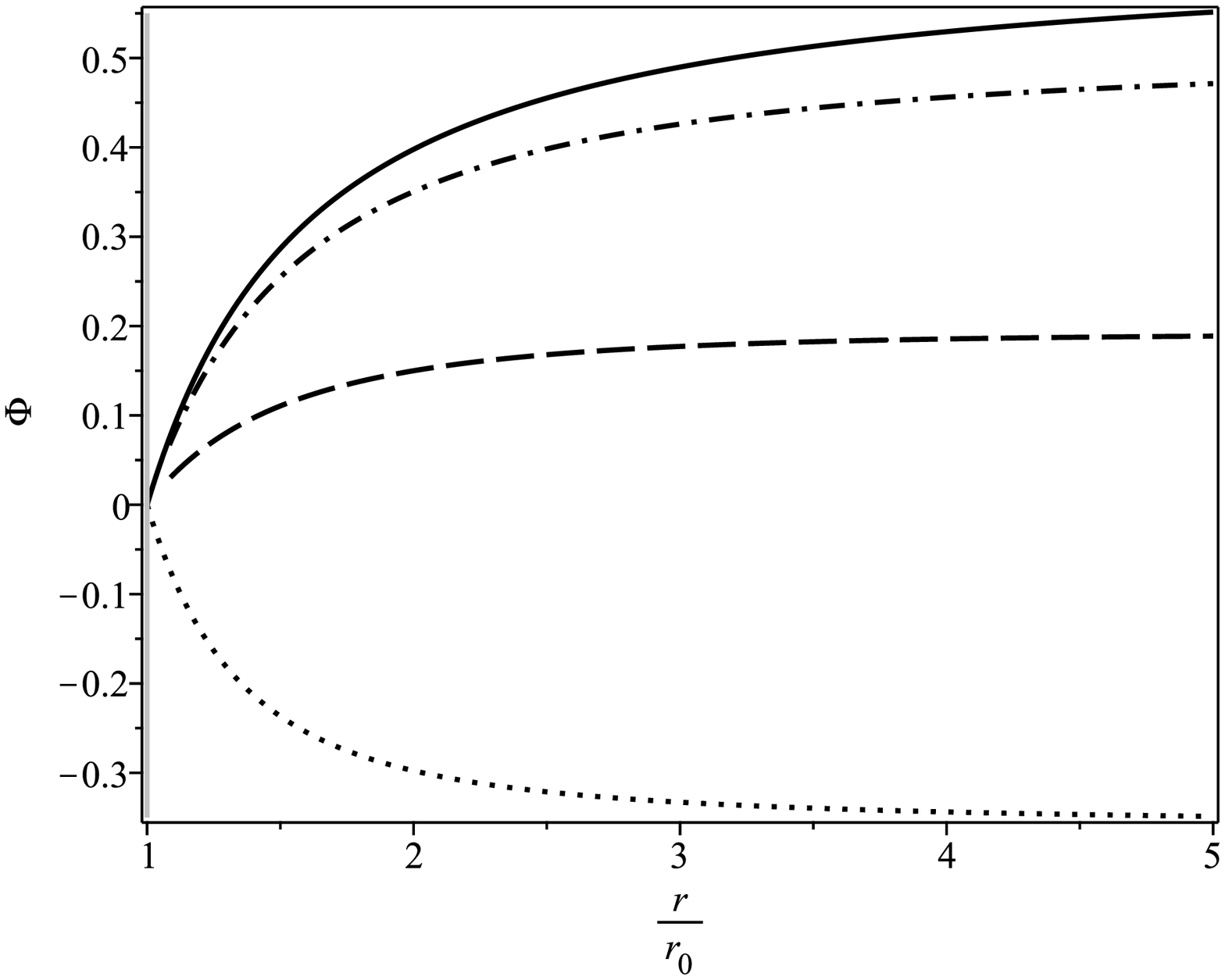}
		\caption{Redshift function, $\Phi(r)$.}
		\label{fig:redcte}
	\end{subfigure}
	\hfill
	\begin{subfigure}{0.48\textwidth}
		\includegraphics[width=\textwidth]{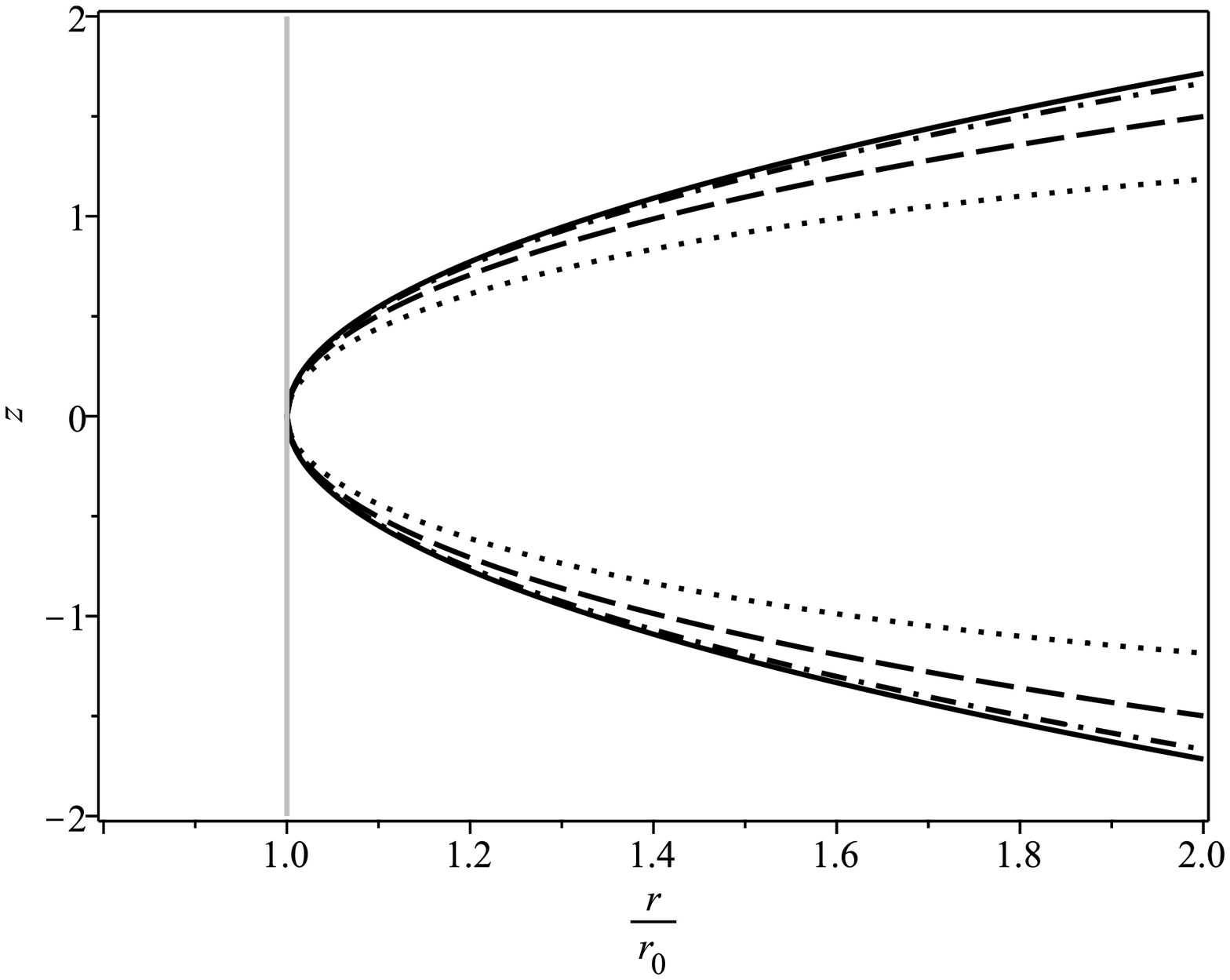}
		\caption{Embedding diagram, $z(r)$.}
		\label{fig:embcte}
	\end{subfigure}
	
	\caption{In all graphs we have $\mu = 0.0$ (solid line), $\mu = 0.1$ (dashdot line), $\mu = 0.5$ (dashed line) and $\mu = 1.0$ (dotted line).}
	\label{fig:cte2}
\end{figure}

An intriguing feature of the global and constant cases is that, since $\Phi'$ is not monotone, i.e. does not have a well-defined sign given the values of $\mu$, it is observed the existence of a critical point in the tangential pressure, see Fig. \ref{fig:tanglobal} and \ref{fig:tangcte}. Thus, if $\Phi_\mu(r)$ characterizes the redshift for a given value of the shielding parameter, and $\mu_c$ characterizes it as a value that exhibits a critical point at $p_t(r)$ then $\Phi_{\mu_c}(r) = 0$ for all $r\geq r_0$, making the Yukawa-Casimir wormhole obey ZTC as long as $b_{\mu_c}(r)$ satisfies the flare-out condition. Also, for $0\leq \mu < \mu_c$ the Wormhole exerts an attractive gravity, but for $\mu > \mu_c$, it becomes repulsive and the redshift turns into a blueshift deviation. This behavior is illustrated in Fig. \ref{fig:redglobal} and \ref{fig:redcte}.

\subsection{Correction in the variable term}

The last proposed correction is to apply the shielding factor in the term with radial dependence of the shape function
\begin{equation}
	b(r) = r_0 - \frac{r_1^2}{r_0} + \frac{r_1^2}{r}e^{-\mu(r - r_0)},
\end{equation}
therefore, this correction ignores the term which is already naturally disregarded in the evaluation of most quantities of physical importance, since they depend on the derivative of the shape function. Thus, it is intuitive to assume that variations in the shielding parameter cause only subtle variations in the quantities of interest.

This correction, as expected, obeys all the conditions that the Casimir shape function already obeyed, in particular
\begin{equation}
	b'(r_0) = -\frac{r_1^2(\mu r_0 + 1)}{r_0^2} < 1.
\end{equation}

Thus, we obtain the energy density that generates such a wormhole
\begin{equation}
	\rho(r) = -\frac{r_1^2}{kr^4}(\mu r + 1)e^{-\mu(r - r_0)},
\end{equation}
in addition, the radial pressure is
\begin{equation}
	p_r(r) = -\frac{r_1^2}{kr^4}(\mu^2r^2 + 3\mu r + 3)e^{-\mu(r - r_0)},
\end{equation}
whose non-homogeneous EoS factor is
\begin{equation}
	\omega(r) = \frac{\mu^2 r^2 + 3\mu r + 3}{\mu r + 1},
\end{equation}
which naturally results in $\omega = 3$ for $\mu = 0$. Note that the only possibility of having $\omega(r_0) = 3$ is having $\mu = 0$, which invalidates the correction, or $r_0 = 0$, which makes the solution unfeasible as a wormhole.

\begin{figure}[!h]
	\centering
	\begin{subfigure}{0.48\textwidth}
		\includegraphics[width=\textwidth]{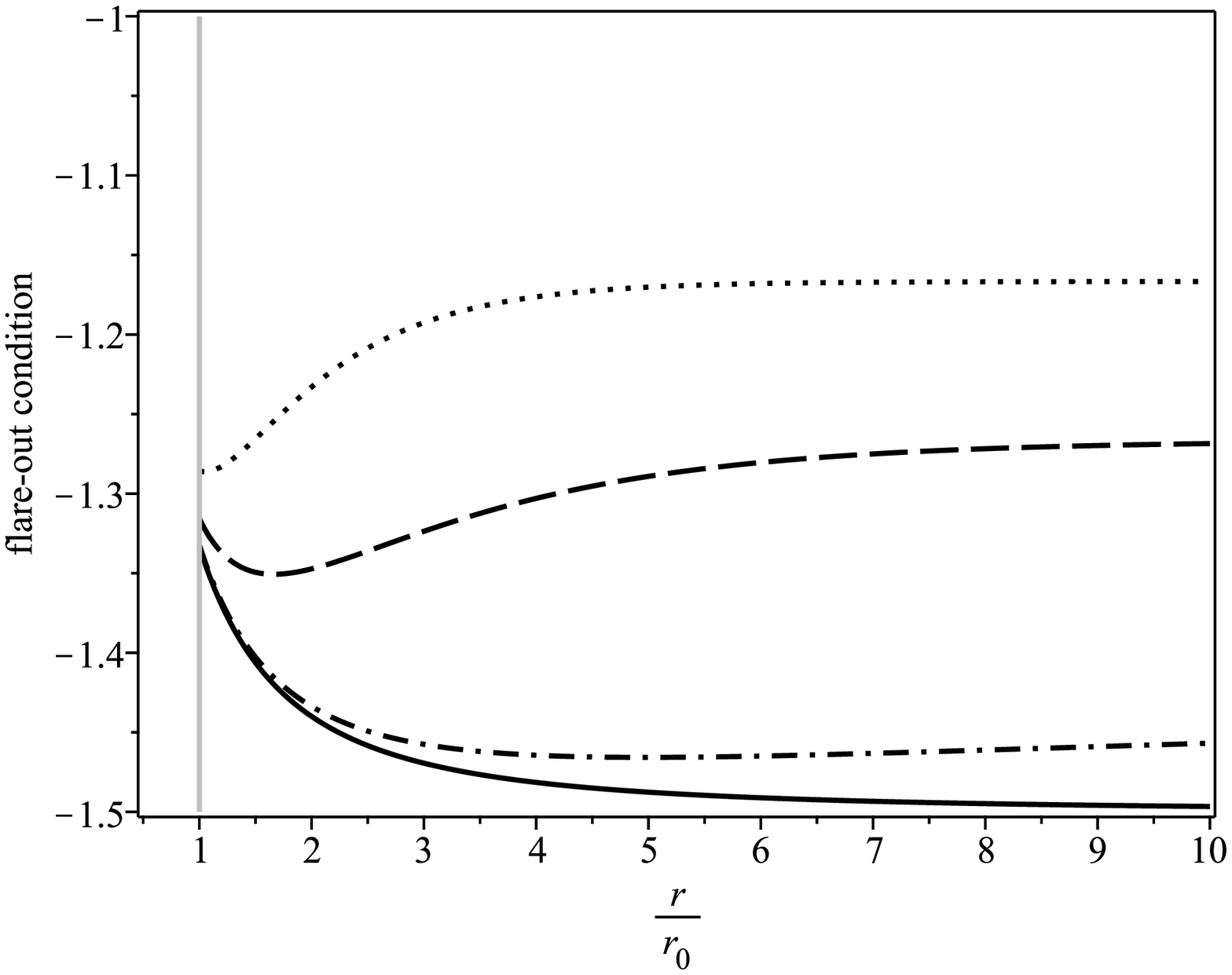}
		\caption{Flare-out condition, $[b'(r)r - b(r)]/b^2(r)$.}
		\label{fig:shapelinear}
	\end{subfigure}
	\hfill
	\begin{subfigure}{0.48\textwidth}
		\includegraphics[width=\textwidth]{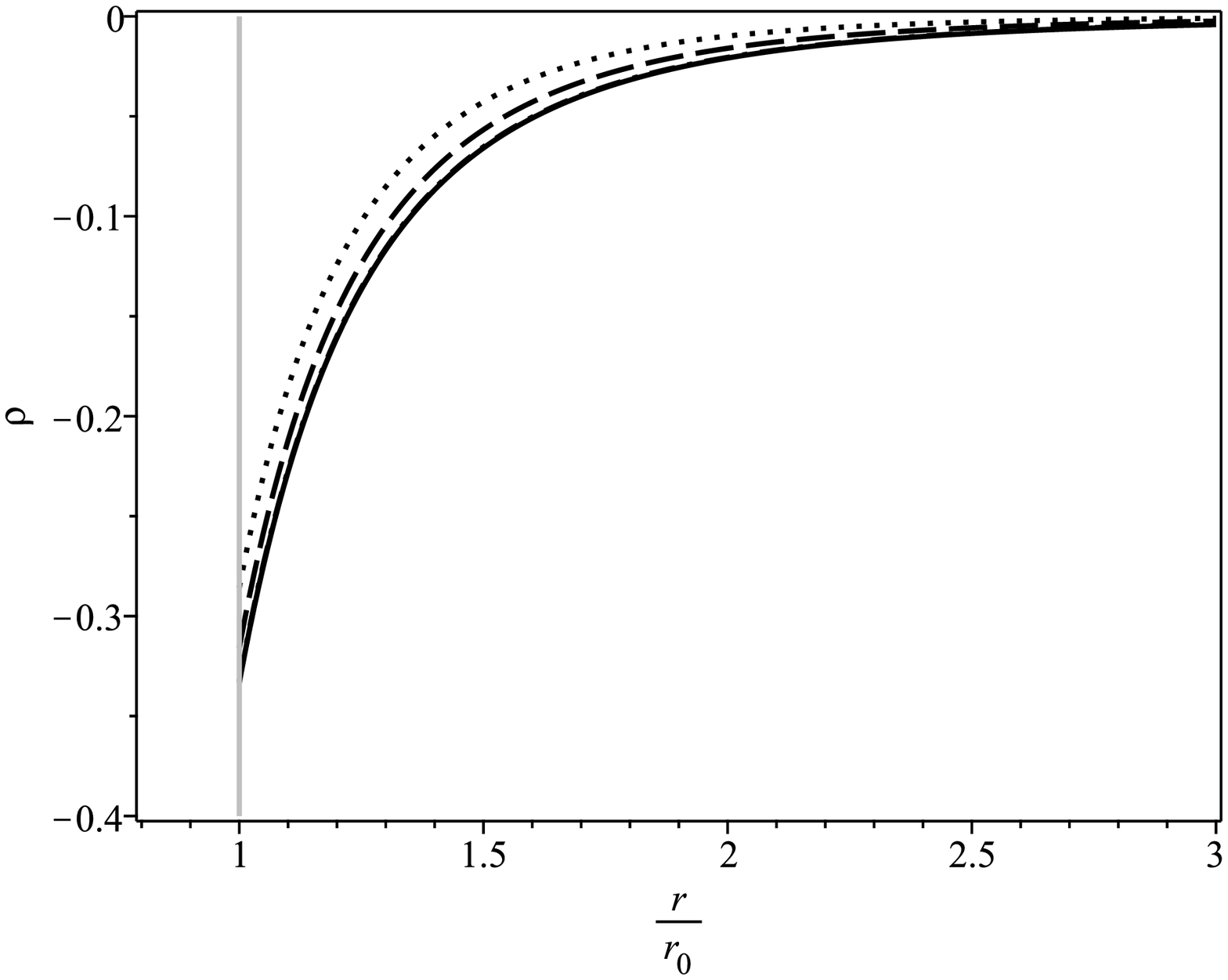}
		\caption{Energy density, $\rho(r)$.}
		\label{fig:energylinear}
	\end{subfigure}
	\hfill
	\begin{subfigure}{0.48\textwidth}
		\includegraphics[width=\textwidth]{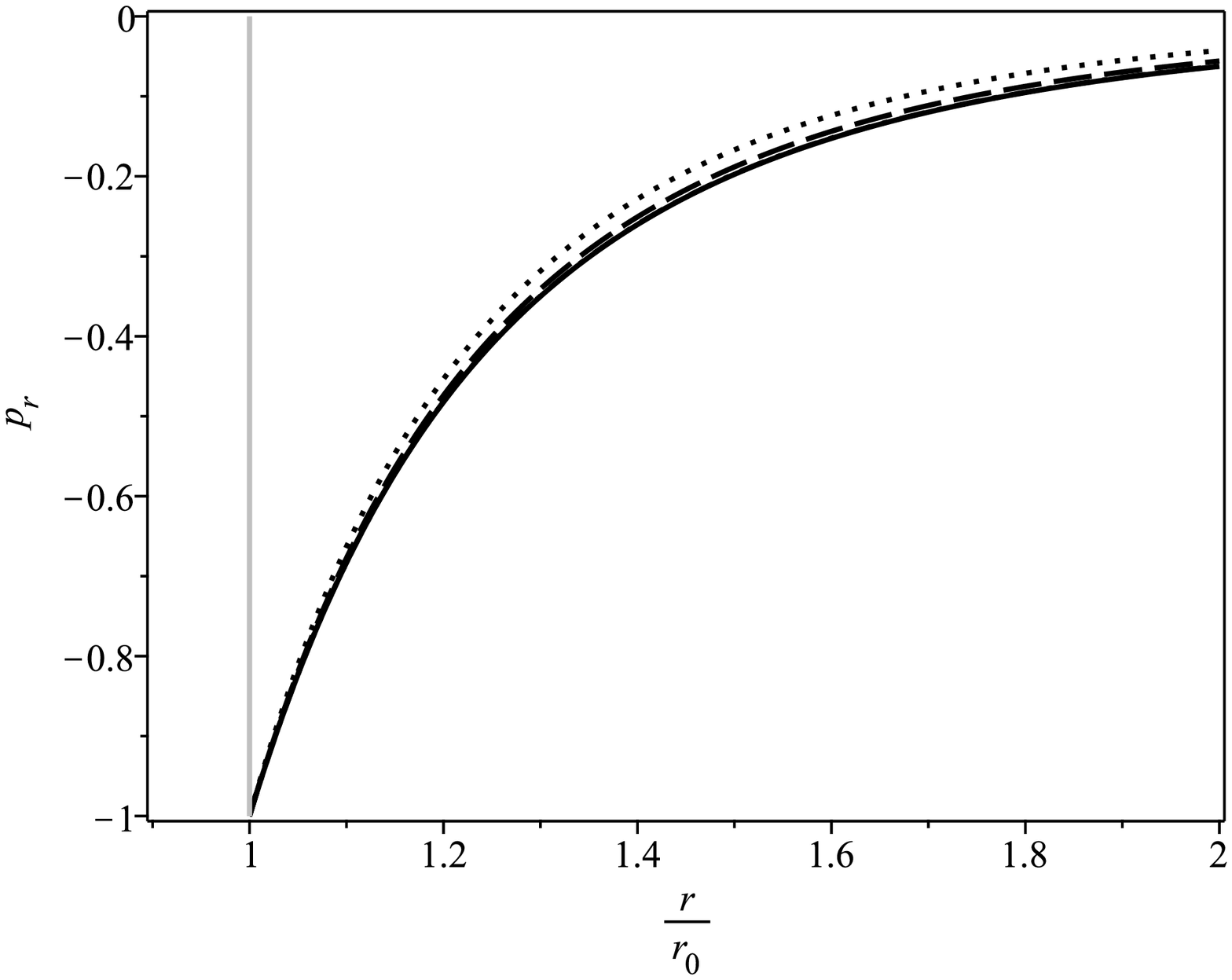}
		\caption{Radial pressure, $p_r(r)$.}
		\label{fig:radiallinear}
	\end{subfigure}
	\hfill
	\begin{subfigure}{0.48\textwidth}
		\includegraphics[width=\textwidth]{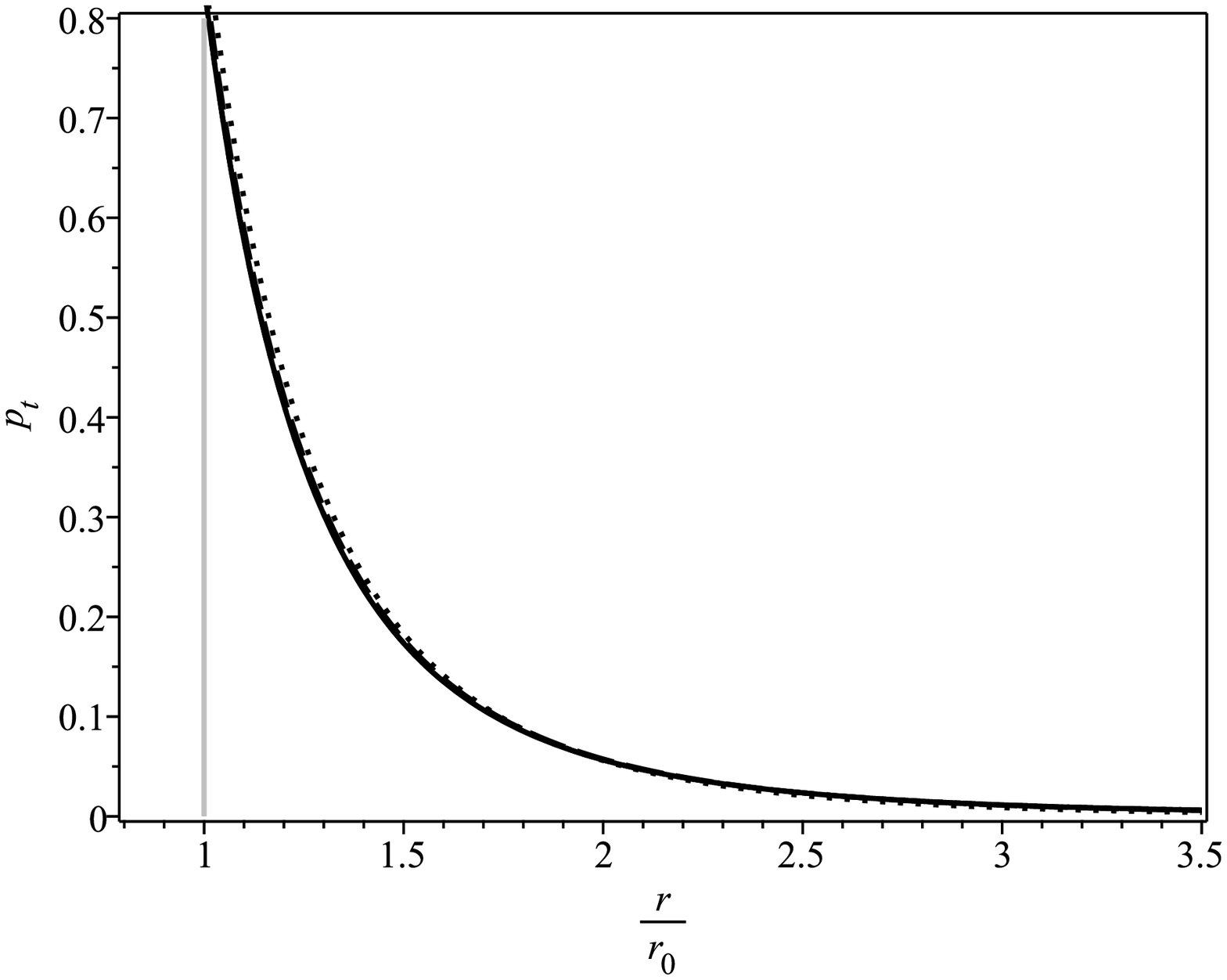}
		\caption{Tangencial pressure, $p_t(r)$.}
		\label{fig:tanglinear}
	\end{subfigure}
	
	\caption{Behavior of the flare-out condition and of the moment-energy tensor components associated with the correction in the variable term. In all graphs we have $\mu = 0.0$ (solid line), $\mu = 0.1$ (dashdot line), $\mu = 0.5$ (dashed line) and $\mu = 1.0$ (dotted line).}
	\label{fig:linear}
\end{figure}

With the second EFE (\ref{eq:EFE2}) we obtain the following fixation between the constants in order to circumvent the existence of horizons in the throat
\begin{equation}
	r_1^2 = \frac{r_0^2}{\mu^2r_0^2 + 3\mu r_0 + 3},
\end{equation}
which results in, by adopting the previously defined radial coordinate scales and redefinitions, the following dimensionless energy
\begin{equation}
	\bar{\rho}(u) = -\frac{1}{u^4}\frac{(\bar{\mu} u + 1)}{(\bar{\mu}^2 + 3\bar{\mu} + 3)}e^{-\bar{\mu}(u - 1)}.
\end{equation}

Analogously, the tangential pressure is obtained via the conservation law, whose result is expressed in Fig. \ref{fig:tanglinear}. In this case, as the derivative of the redshift function is monotonic, it is not possible to obtain solutions without tidal force.

\begin{figure}[!h]
	\centering
	\begin{subfigure}{0.48\textwidth}
		\includegraphics[width=\textwidth]{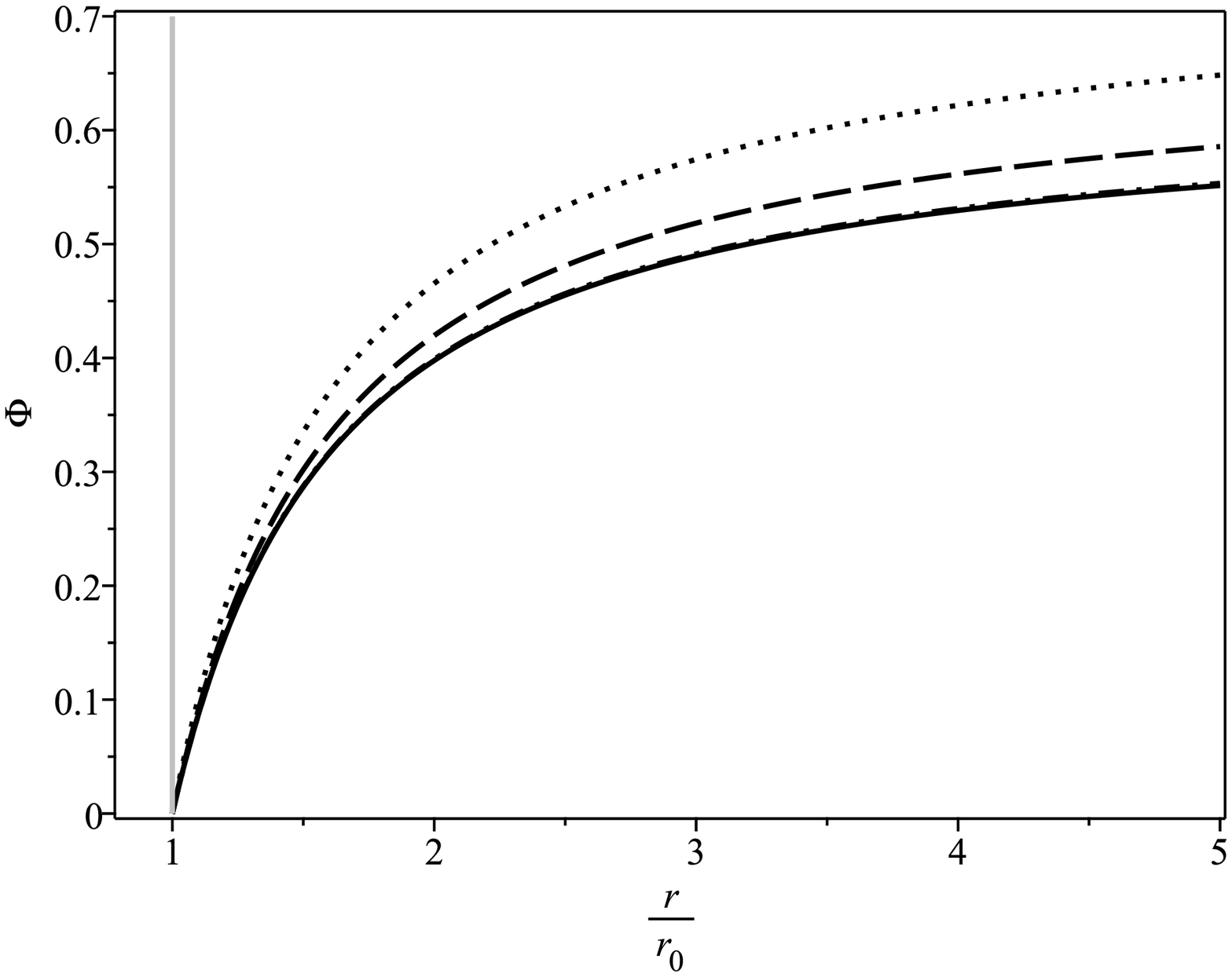}
		\caption{Redshift function, $\Phi(r)$.}
		\label{fig:redlinear}
	\end{subfigure}
	\hfill
	\begin{subfigure}{0.48\textwidth}
		\includegraphics[width=\textwidth]{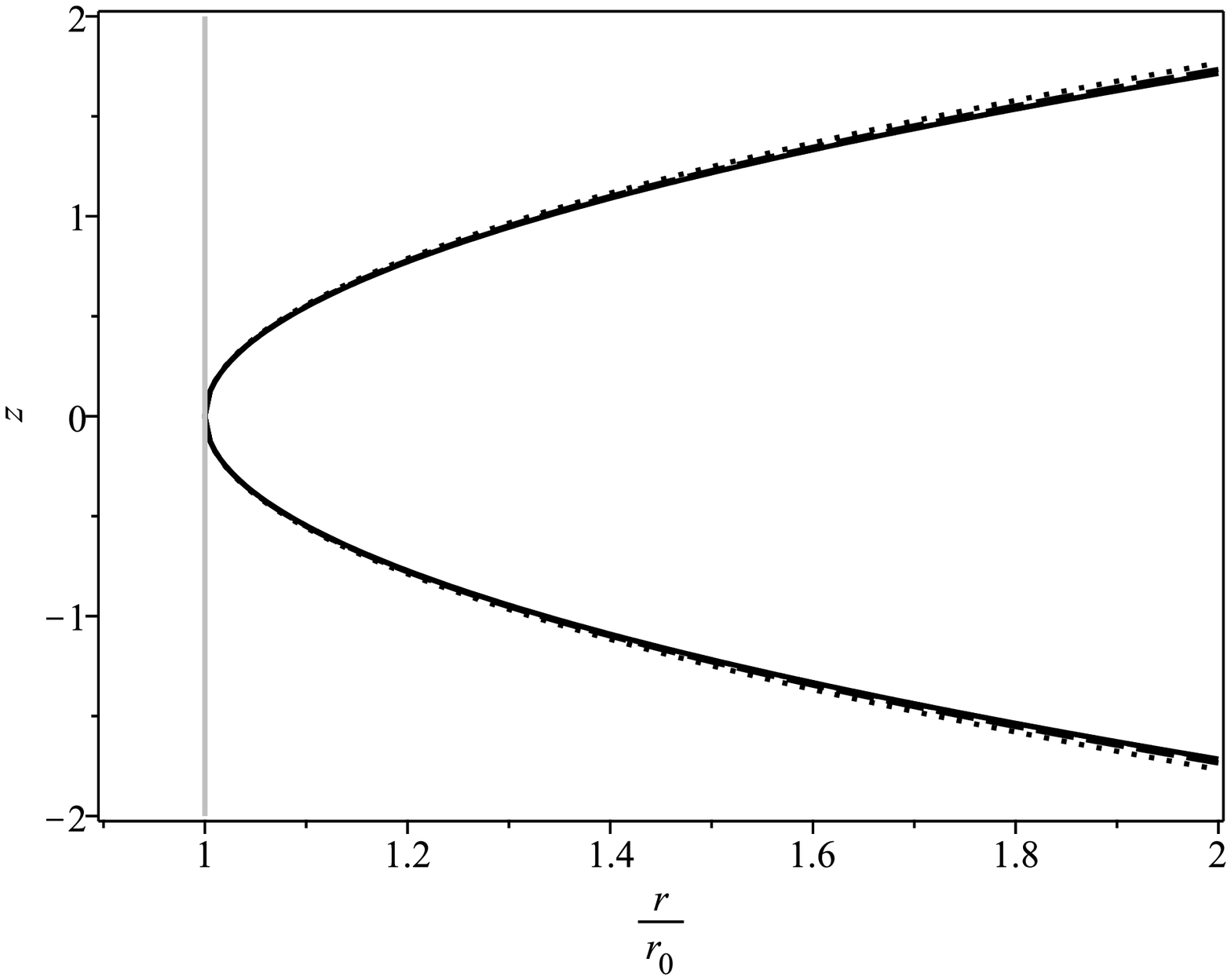}
		\caption{Embedding diagram, $z(r)$.}
		\label{fig:emblinear}
	\end{subfigure}
	
	\caption{In all graphs we have $\mu = 0.0$ (solid line), $\mu = 0.1$ (dashdot line), $\mu = 0.5$ (dashed line) and $\mu = 1.0$ (dotted line).}
	\label{fig:linear2}
\end{figure}

\subsection{Correction for small parameters}

Although it is not possible to obtain a complete analytical solution for the redshift function in the cases considered, a complementary analysis is to compare the three Casimir wormhole generalization models for small values of the parameter $\mu$, thus the local exponential terms in $\Phi'$ can be expressed via Taylor series
\begin{equation}
    e^{-\mu(r - r_0)} = 1 - \mu(r - r_0) + \frac{1}{2}\mu^2(r - r_0)^2 + \cdot\cdot\cdot,
\end{equation}
so a first-order correction would be to take all terms of the type $\mu^j$ with $j \leq 1$, second-order with $j \leq 2$, and so on.

\subsubsection{First order correction}

Let's consider first-order corrections to the redshift function. In that case, we have
\begin{subequations}
	\begin{eqnarray}
		\Phi'_{\textnormal{global}}(r) &\approx& \frac{r_0}{(3r + r_0)r} - \frac{6(r + r_0)r_0}{(3r + r_0)^2}\mu,\\
		\Phi'_{\textnormal{constant}}(r) &=& \Phi'_{\textnormal{global}}(r),\\
		\Phi'_{\textnormal{radial}}(r) &\approx& \frac{r_0}{(3r + r_0)r},
	\end{eqnarray}
\end{subequations}
thus, the first-order corrected redshift functions of $\mu$ are given by,
\begin{subequations}
	\begin{eqnarray}
		\Phi_{\textnormal{global}}(r) &\approx& \frac{(\mu r_0+\frac{3}{2})(-6r-2r_0)\ln(\frac{3r}{4}+\frac{r_0}{4})+6\mu r_0\left(r+\frac{r_0}{3}\right)\ln(r_0)+(3r_0+9r)\ln(r)-3\mu r_0(r-r_0)}{3r_0+9r},\nonumber\\
		\Phi_{\textnormal{radial}}(r) &\approx& \ln\left(\frac{4r}{3r + r_0}\right)\nonumber.
	\end{eqnarray}
\end{subequations}

Note that, in the radial case, the solution is basically the redshift of the Casimir wormholes, but with a subtle change in the numerator of the logarithm argument compared to Eq. \eqref{eq:redcasimir}. This change occurs by fixing $\Phi(r_0)$, so that the redshift is set to less than an additive constant, with no physical change. On the other hand, the global and constant cases behave the same way for first-order changes.

\subsubsection{Second order correction}

\begin{figure}[!h]
	\centering
	\includegraphics[width=0.7\linewidth]{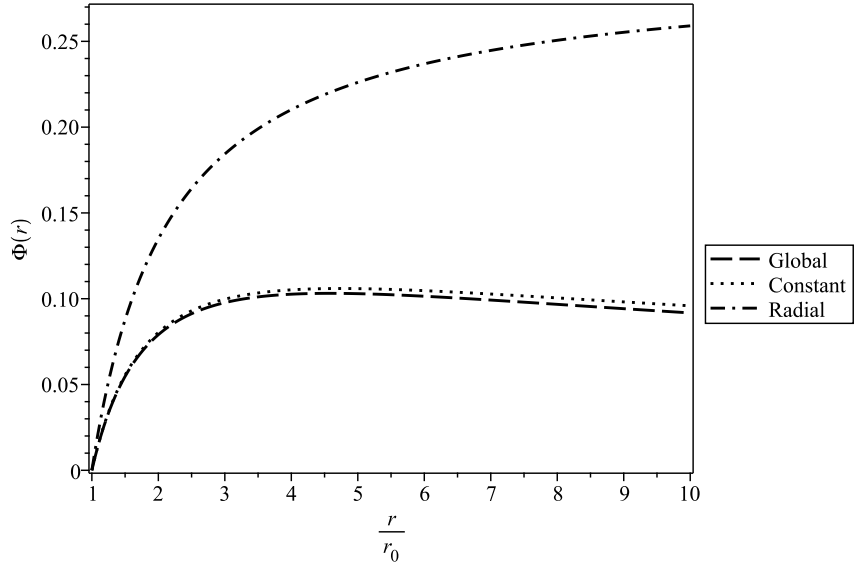}
	\caption{Behavior of the second-order modified redshift function on the parameter $\mu$ for Yukawa-Casimir wormholes.}
	\label{fig:second-order}
\end{figure}

As for first order the global and constant cases result in the same and the radial case becomes independent of $\mu$ it is suggestive to consider corrections in second order. So,
\begin{subequations}
	\begin{eqnarray}
		\Phi'_{\textnormal{global}}(r) &\approx& \frac{r_0}{(3 r+r_0)r}-\frac{6 (r+r_0) r_0}{(3 r+r_0)^{2}}\mu+\frac{3 (3 r^{3}-4 r^{2} r_0-15 r_0^{2} r-8 r_0^{3}) r_0}{2 (3 r+r_0)^{3}}\mu^{2},\\
		\Phi'_{\textnormal{constant}}(r) &\approx& \frac{r_0}{(3 r+r_0) r}-\frac{6 (r+r_0) r_0}{(3 r+r_0)^{2}}\mu+\frac{3 (3 r-7 r_0) r_0 (r+r_0)^{2}}{2 (3 r+r_0)^{3}}\mu^{2},\\
		\Phi'_{\textnormal{radial}}(r) &\approx& \frac{r_0}{(3r + r_0)r} + \frac{3r_0^2(r + r_0)}{2(3r + r_0)^2}\mu^2,
	\end{eqnarray}
\end{subequations}
thus, the second-order corrected redshift functions of $\mu$ are given by,
\begin{subequations}
	\begin{eqnarray}
		\Phi_{\textnormal{global}}(r) &\approx& \frac{1}{36 (3 r+r_0)^{2}}\left[\left(r+\frac{r_0}{3}\right)^{2} (-126 \mu^{2} r_0^{2}-216 \mu r_0-324) \ln\left(\frac{3 r}{4}+\frac{r_0}{4}\right)\right.\nonumber\\&&\left.+126 r_0 \left(r+\frac{r_0}{3}\right)^{2} \left(\mu r_0+\frac{12}{7}\right) \mu \ln(r_0) + 324 \left(r+\frac{r_0}{3}\right)^{2} \ln(r)\right.\nonumber\\&&\left.+54 r_0 \left(-\frac{25 \mu r_0^{2}}{18}+\left(-\frac{5 \mu r}{2}-\frac{2}{3}\right) r_0+\mu r^{2}-2 r\right) \mu (r-r_0)\right],\\
		\Phi_{\textnormal{constant}}(r) &\approx& \frac{1}{18 (3 r+r_0)^{2}}\left[\left(r+\frac{r_0}{3}\right)^{2} (-36 \mu^{2} r_0^{2}-108 \mu r_0-162) \ln\left(\frac{3 r}{4}+\frac{r_0}{4}\right)\right.\nonumber\\&&\left.+36 r_0 \left(r+\frac{r_0}{3}\right)^{2} \mu (\mu r_0+3) \ln(r_0)+162 \left(r+\frac{r_0}{3}\right)^{2} \ln(r)\right.\nonumber\\&&\left.+27 r_0 \left(-\frac{11 \mu r_0^{2}}{9}+\left(-2 r \mu-\frac{2}{3}\right) r_0+\mu r^{2}-2 r\right) (r-r_0) \mu\right],\\
		\Phi_{\textnormal{radial}}(r) &\approx& \frac{1}{36r+12r_0}\left[(\mu^{2}r_0^{2}-6)(6r+2r_0)\ln(\frac{3r}{4}+\frac{r_0}{4})-6\mu^{2}(r+\frac{r_0}{3})r_0^{2}\ln(r_0)+(36r+12r_0)\ln(r)\right.\nonumber\\&&\left.+3\mu^{2}r_0^{2}(r-r_0)\right],
	\end{eqnarray}
\end{subequations}
the behavior of these solutions as a function of $r/r_0$ is plotted in Fig. \ref{fig:second-order}. In this order of approximation, the global and radial cases differ from each other, although they are subtle.

\section{Conclusion}

\begin{figure}[!h]
	\centering
	\begin{subfigure}{0.48\textwidth}
		\includegraphics[width=\textwidth]{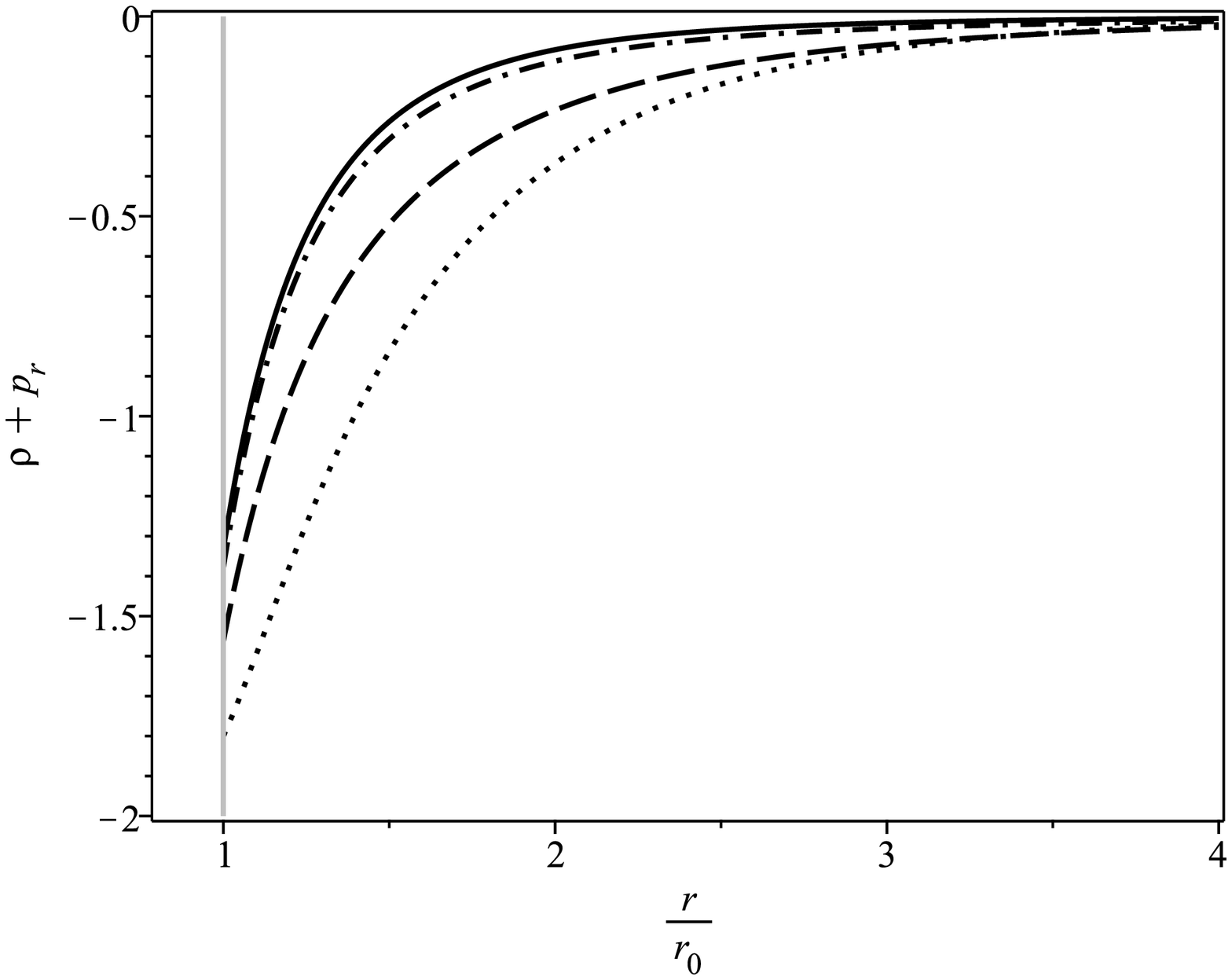}
		\caption{Global correction.}
		\label{fig:qwec1}
	\end{subfigure}
	\hfill
	\begin{subfigure}{0.48\textwidth}
		\includegraphics[width=\textwidth]{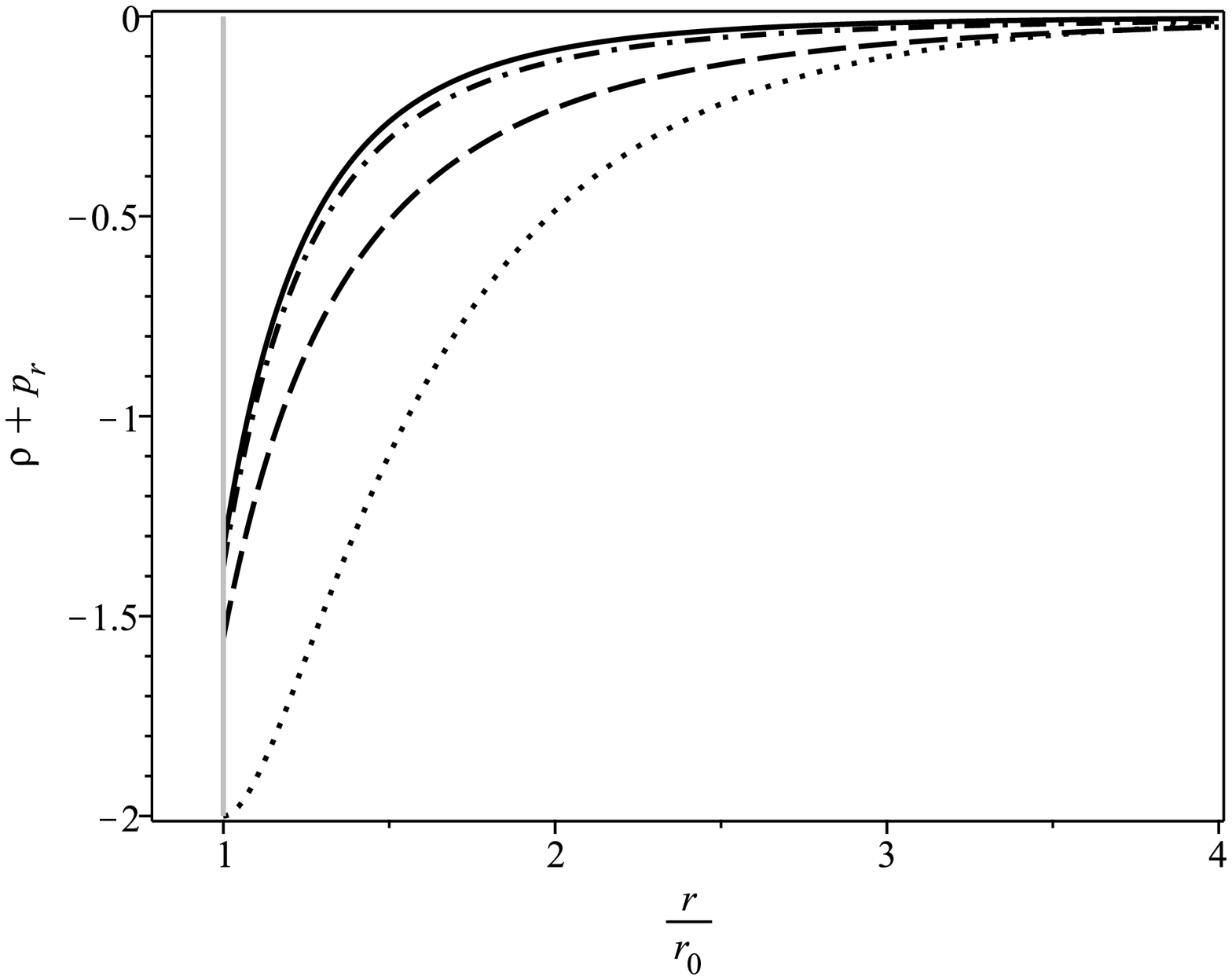}
		\caption{Constant correction.}
		\label{fig:qwec2}
	\end{subfigure}
	\hfill
	\begin{subfigure}{0.48\textwidth}
		\includegraphics[width=\textwidth]{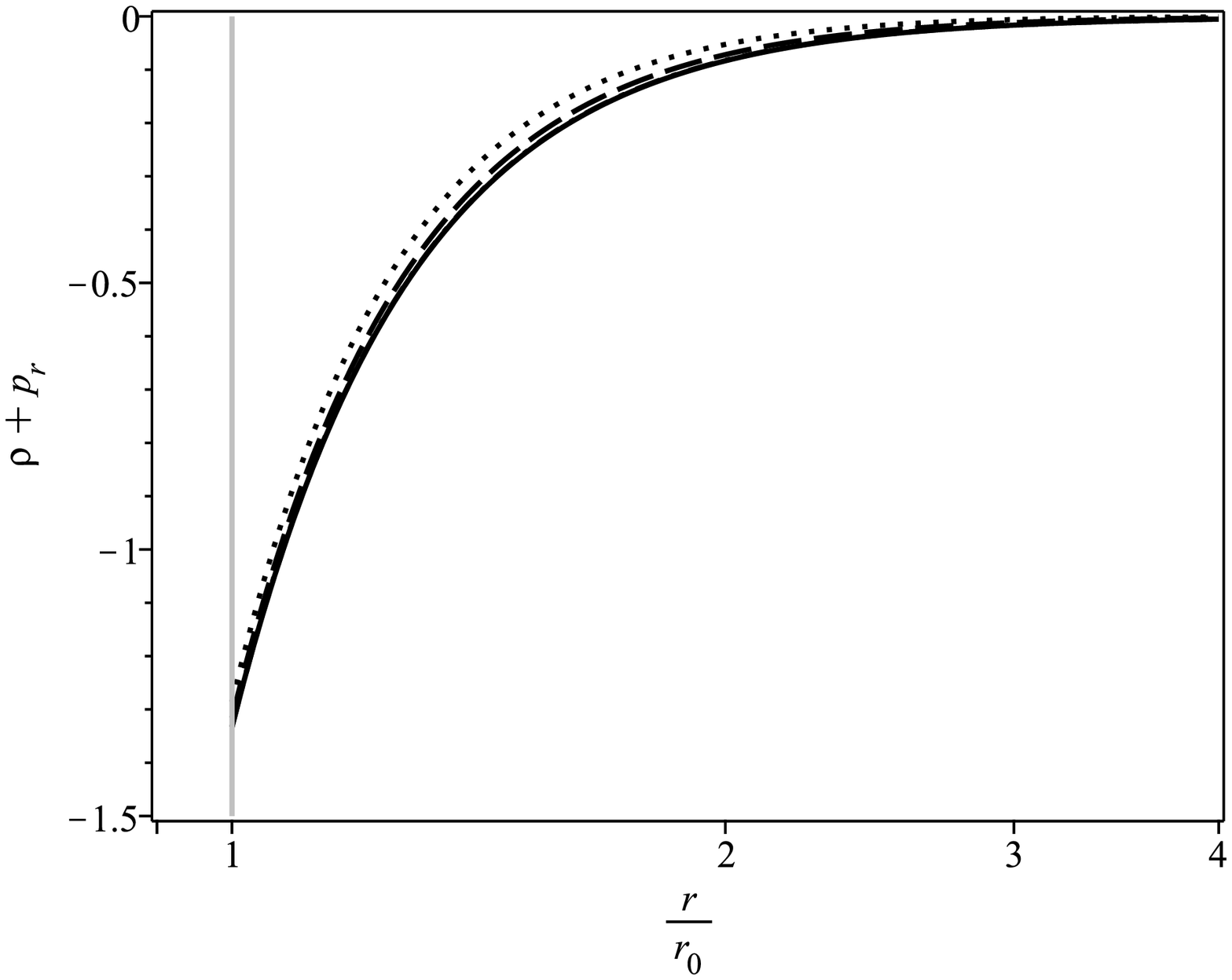}
		\caption{Linear correction.}
		\label{fig:qwec3}
	\end{subfigure}	
	\caption{Quantum Weak Energy Condition (QWEC) check. In all graphs we have $\mu = 0.0$ (solid line), $\mu = 0.1$ (dashdot line), $\mu = 0.5$ (dashed line) and $\mu = 1.0$ (dotted line).}
	\label{fig:qwec}
\end{figure}

In this work, we propose a more consistent way to generalize Casimir wormholes with Yukawa-type shielding terms in shape functions in the three approaches initially proposed by Garattini \cite{Garattini:2021kca}. In possession of the results obtained, the Yukawa-Casimir wormholes, in all considered cases, satisfy the so-called QWEC (Quantum Weak Energy Condition) \cite{Garattini:2019ivd}
\begin{equation}
	\rho(r) + p_r(r) = -f(r); \hspace{1cm} f(r) > 0 \ \forall \ r \in[r_0,\infty),
\end{equation}
as visualized in Fig. (\ref{fig:qwec}). In the correction in the variable term, due to the subtle discrepancies between the curves, a logarithmic scale was considered in the independent variable $r/r_0$.

Furthermore, only in the correction of the constant term do restrictions appear on the value of the shielding constant, since $\mu = 1.0$ violates the flare-out condition. Another characteristic of this methodology was a natural achievement, in the global and constant cases, of being possible not only to construct wormholes without tidal force but also to obtain cases in which the generated force is repulsive, characterizing a blueshift deviation.

In all cases, the complete solution of the redshift function cannot be obtained analytically due to local exponential terms in the integrand. Therefore, two approaches were taken. Initially, numerical integration was performed for the variable $r/r_0$ and setting $\mu = 0.1$, which were reproduced in Figs. (\ref{fig:redglobal}), (\ref{fig:redcte}) and (\ref{fig:redlinear}). Finally, an analysis was performed for small values of the shielding parameter, which in the first order revealed that $\Phi_{\textnormal{global}}(r) = \Phi_{\textnormal{constant}}(r)$, while for the radial case the redshift function does not explicitly depend on the shielding parameter. On the other hand, in second-order, it was found that the global and constant cases become slightly different.

\section*{Acknowledgement}
The authors would like to thanks the financial support provided by the Coordenação de Aperfeiçoamento de Pessoal de Nível Superior (CAPES) and by the Conselho Nacional de Desenvolvimento Científico e Tecnológico (CNPq) through Universal 315568/2021-6. We also acknowledge Fundação Cearense de Apoio ao Desenvolvimento Científico e Tecnológico (FUNCAP) through PRONEM PNE0112- 00085.01.00/16.


\begin{thebibliography}{00}

\bibitem{Schwarzschild:1916uq}
K.~Schwarzschild,
Sitzungsber. Preuss. Akad. Wiss. Berlin (Math. Phys.) \textbf{1916}, 189-196 (1916)
[arXiv:physics/9905030 [physics]].

\bibitem{Akiyama:2019cqa}
K.~Akiyama \textit{et al.} [Event Horizon Telescope],
Astrophys. J. Lett. \textbf{875}, L1 (2019)
doi:10.3847/2041-8213/ab0ec7
[arXiv:1906.11238 [astro-ph.GA]].

\bibitem{Akiyama:2021tfw}
K.~Akiyama, J.~C.~Algaba, A.~Alberdi, W.~Alef, R.~Anantua, K.~Asada, R.~Azulay, A.~K.~Baczko, D.~Ball and M.~Balokovi\'c, \textit{et al.}
Astrophys. J. Lett. \textbf{910}, no.1, L13 (2021)
doi:10.3847/2041-8213/abe4de
[arXiv:2105.01173 [astro-ph.HE]].

\bibitem{Hassan:2022ibc}
Z.~Hassan, S.~Ghosh, P.~K.~Sahoo and V.~S.~H.~Rao,
[arXiv:2209.02704 [gr-qc]].

\bibitem{Shinkai:2002gv}
H.~A.~Shinkai and S.~A.~Hayward,
Phys. Rev. D \textbf{66}, 044005 (2002)
doi:10.1103/PhysRevD.66.044005
[arXiv:gr-qc/0205041 [gr-qc]].

\bibitem{Maldacena:2020sxe}
J.~Maldacena and A.~Milekhin,
Phys. Rev. D \textbf{103}, no.6, 066007 (2021)
doi:10.1103/PhysRevD.103.066007
[arXiv:2008.06618 [hep-th]].

\bibitem{Gao:2016bin}
P.~Gao, D.~L.~Jafferis and A.~C.~Wall,
JHEP \textbf{12}, 151 (2017)
doi:10.1007/JHEP12(2017)151
[arXiv:1608.05687 [hep-th]].

\bibitem{Casimir:1948dh}
H.~B.~G.~Casimir,
Indag. Math. \textbf{10}, 261-263 (1948)

\bibitem{Avino:2019fdq}
S.~Avino, E.~Calloni, S.~Caprara, M.~De Laurentis, R.~De Rosa, T.~Di Girolamo, L.~Errico, G.~Gagliardi, M.~Grilli and V.~Mangano, \textit{et al.}
MDPI Physics \textbf{2}, no.1, 1-13 (2020)
doi:10.3390/physics2010001

\bibitem{Garattini:2019ivd}
R.~Garattini,
Eur. Phys. J. C \textbf{79}, no.11, 951 (2019)
doi:10.1140/epjc/s10052-019-7468-y
[arXiv:1907.03623 [gr-qc]].

\bibitem{Oliveira:2021ypz}
P.~H.~F.~Oliveira, G.~Alencar, I.~C.~Jardim and R.~R.~Landim,
Mod. Phys. Lett. A \textbf{37}, no.15, 2250090 (2022)
doi:10.1142/S0217732322500900
[arXiv:2107.00605 [hep-th]].

\bibitem{Garattini:2021kca}
R.~Garattini,
Eur. Phys. J. C \textbf{81}, no.9, 824 (2021)
doi:10.1140/epjc/s10052-021-09634-3
[arXiv:2107.09276 [gr-qc]].

\bibitem{Morris:1988cz}
M.~S.~Morris and K.~S.~Thorne,
Am. J. Phys. \textbf{56}, 395-412 (1988)
doi:10.1119/1.15620

\bibitem{Alnes:2006pa}
H.~Alnes, F.~Ravndal, I.~K.~Wehus and K.~Olaussen,
Phys. Rev. D \textbf{74}, 105017 (2006)
doi:10.1103/PhysRevD.74.105017
[arXiv:quant-ph/0607081 [quant-ph]].




\end{thebibliography}

\end{document}